\documentclass[10pt,sigconf,letterpaper,nonacm,screen]{acmart}

\PassOptionsToPackage{dvipsnames}{xcolor}

\PassOptionsToPackage{colorlinks=true,linkcolor=RubineRed,breaklinks=True,citecolor=RubineRed}{xcolor}

\usepackage{enumitem}
\setlist[itemize]{noitemsep, topsep=0pt}

\usepackage[most]{tcolorbox}
\usepackage{adjustbox}
\usepackage[linesnumbered,algo2e,ruled]{algorithm2e}
\usepackage{multirow}
\usepackage{booktabs,subcaption,dcolumn}
\usepackage{tikz}
\usepackage{enumitem}
\usepackage{tabularx}
\usepackage{MnSymbol}
\usepackage{setspace}

\usepackage{pifont}
\usepackage{svg}
\usepackage{soul}
\usepackage{amsmath}
\usepackage{dirtytalk}
\usepackage{subcaption}
\usepackage{graphicx}
\usepackage{setspace}
\usepackage[font=small]{caption}
\usepackage{booktabs} 
\usepackage[titletoc,toc,title]{appendix}
\usepackage{svg}
\usepackage{amsfonts}
\usepackage{xurl} 

\usepackage{diagbox}
\usepackage{indentfirst}

\usepackage{CJKutf8}
\usepackage{amsmath}
\usepackage{amsfonts} 

\usepackage{colortbl}

\newcounter{emailbox}

\setcopyright{none}

\AtBeginDocument{%
  \providecommand\BibTeX{{%
    \normalfont B\kern-0.5em{\scshape i\kern-0.25em b}\kern-0.8em\TeX}}}

\setcopyright{acmcopyright}
\copyrightyear{2023}
\acmYear{2023}
\acmDOI{}

\newtcolorbox{emailBox}[2][]{%
  colback=gray!5!white,
  colframe=gray!50!white,
  boxrule=0.5pt,
  sharp corners,
  title={\textbf{[Subject] #2}},    
  fonttitle=\small\sffamily,
  after lower={},
  nobeforeafter,
  breakable,
  left=2pt, right=2pt, top=0pt, bottom=0pt,
  before upper=\refstepcounter{emailbox}, 
  #1
}
\newcolumntype{?}{!{\vrule width 1.5pt}}

\newcommand{\anonym}[0]{{\color{red}***}}

\newcommand{\textbox}[1]{
    \noindent\fbox{%
        \parbox{0.97\columnwidth}{%
            {#1}
        }%
    }
}

\newtcolorbox{cooltextbox}[1][]{%
    colback=black!5,
    colframe=black!5,
    notitle,
    sharp corners,
    borderline west={0pt}{0pt}{red!80!black},
    enhanced,
    breakable,
    left=0pt,
    right=0pt,
    top=0pt,
    bottom=0pt
    }

\newtcolorbox{softtextbox}[1][]{%
    colback=white!5,
    colframe=black!15,
    notitle,
    sharp corners,
    enhanced,
    breakable,
    left=0pt,
    right=0pt,
    top=0pt,
    bottom=0pt
    }

\newcommand\smamath[1]{{\small $#1$}}

\tcbuselibrary{theorems, skins, breakable}

\newtcbtheorem[auto counter]{excerpt}{Excerpt from Comments \#\!\!}{%
  enhanced,
  breakable,
  colback=yellow!20,           
  colframe=black!20,       
  boxrule=0.4pt,
  sharp corners,
  fontupper=\sffamily\scriptsize,     
  fonttitle=\small,
  before skip=6pt, after skip=6pt,
}{stmt}
\begin{document}
\title{X-raying the arXiv: A Large-Scale Analysis of arXiv Submissions' Source Files}


\author{Giovanni Apruzzese}
\orcid{0000-0002-6890-9611}
\affiliation{%
  \institution{University of Liechtenstein}
  \city{Vaduz}
  \country{Liechtenstein}}
\affiliation{%
  \institution{Reykjavik University}
  \city{Reykjavik}
  \country{Iceland}}
\email{giovanni.apruzzese@uni.li}

\author{Aurore Fass}
\orcid{0000-0001-6611-4447}
\affiliation{%
  \institution{Inria Centre at Université Côte d’Azur}
  \city{Sophia Antipolis}
  \country{France}}
\affiliation{%
  \institution{CISPA Helmholtz Center for Information Security}
  \city{Saarbrücken}
  \country{Germany}}
\email{aurore.fass@inria.fr}

\begin{abstract}
arXiv is the largest open-access repository for scientific literature. When submitting a paper, authors upload the manuscript’s source files, from which the final PDF is compiled. These source files are also publicly downloadable, potentially exposing data unrelated to the published paper---such as figures, documents, or comments---that may unintentionally reveal confidential information or simply waste storage space. We thus ask ourselves: ``\textit{What can be found within the source files of arXiv submissions?}''

We present a longitudinal analysis of $\approx$600,000 submissions appeared on arXiv between 2015--2025. For each submission, we examine the uploaded source files to quantify and characterize data not required for producing the respective PDF. On average, 27\% of the data in each submission are unnecessary, totaling >580 GB of redundant content across our dataset. Qualitative inspection reveals the presence of offensive/inappropriate text (e.g., ``\textsf{WTF does this mean?}'') and experimental details that could disclose ongoing research. We have contacted arXiv's leadership team, as well as the authors of affected papers to alert them of these issues. Finally, we propose recommendations and an automated tool to detect and analyze arXiv submissions residual data at scale, aiming to improve data hygiene in the arXiv's ecosystem.
\end{abstract}

\settopmatter{printfolios=true}

\maketitle

\section{Introduction}
\label{sec:introduction}
\noindent
arXiv represents a cornerstone for modern research, enabling the fast and free exchange (upload and download) of ``preprints'' used to disseminate the most recent scientific discoveries~\cite{bagchi2025effects}.
As of August 2025, over 60M connections are made daily to arXiv~\cite{arxivstats}. Its databases encompass over 2.8M submissions, whose overall number of downloads exceeds 3.3 billions~\cite{arxivstats}. Just in Q1 2025, over 65k papers have been submitted to arXiv, spanning a variety of domains---of which computer science detains the lion's share~\cite{arxivstats,lin2020many}.

\begin{figure}[!t]
    \centering
    \frame{\includegraphics[width=1\linewidth]{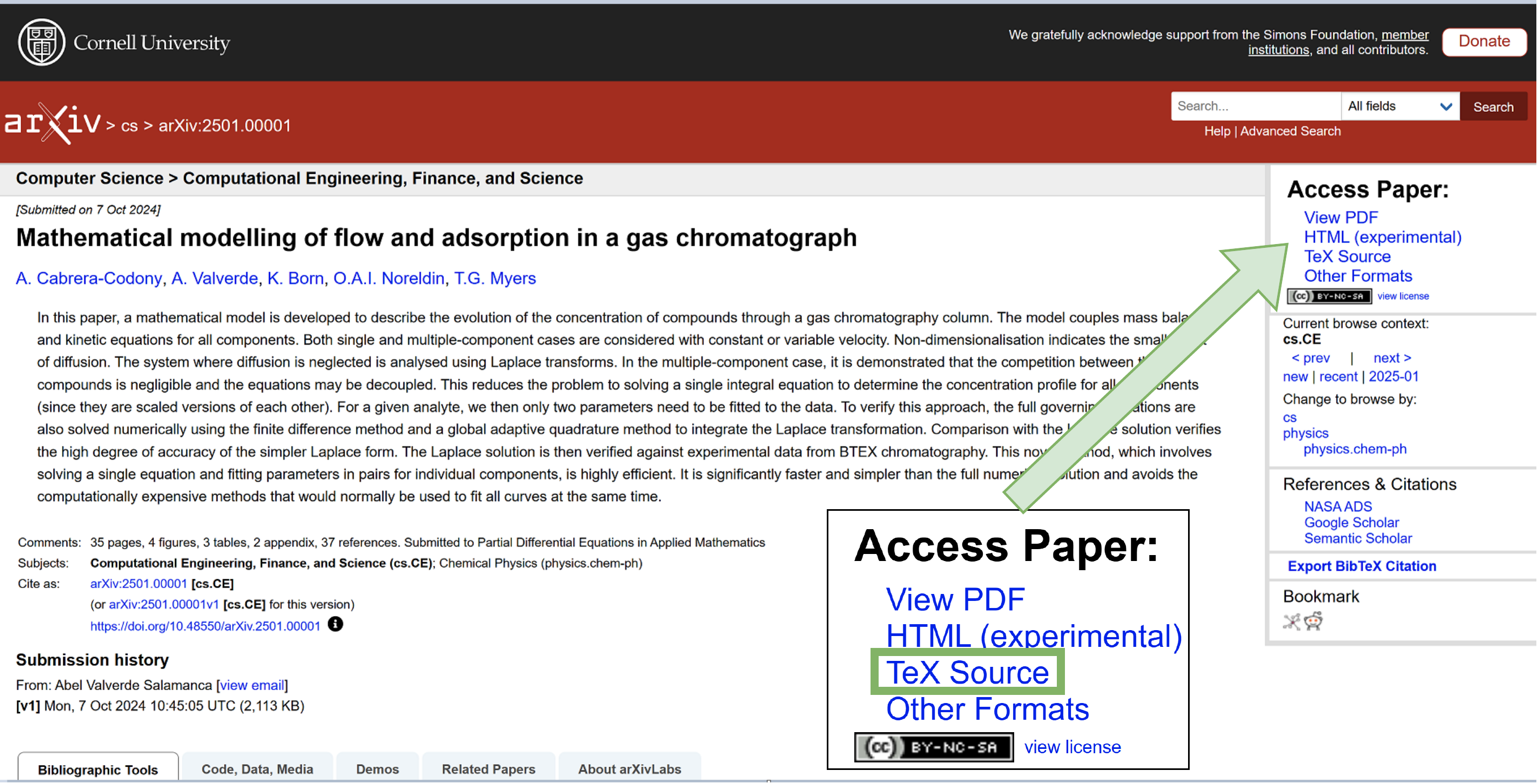}}
    \vspace{-6mm}
    \caption{\textbf{An arXiv submission.} \textmd{Clicking on ``{\footnotesize \TeX{}} Source'' allows anyone to download the source files of this paper.}}
    \label{fig:arxiv_home}
    \vspace{-3mm}
\end{figure}

Uploading a preprint on arXiv is a simple process: an author, potentially after having received an endorsement by another arXiv user, can either {\small \textit{(a)}}~submit a paper as a PDF---but such an option is rarely used (we found that only 10\% of submissions are made in this way) and is not encouraged by arXiv~\cite{arxiv_whypdf}; or {\small \textit{(b)}}~submit the source {\footnotesize \TeX{}} files of a paper, which are then rendered as PDF by arXiv itself---this is the way recommended by arXiv~\cite{arxiv_whytex} and used by most submissions.

arXiv is rooted on the concept of \textit{openness}~\cite{arxiv2024report}. For instance, the web page of a submission provides a history that lists all the previous ``versions'' of such a submission. All such versions are freely retrievable (in PDF format); even submissions that have been withdrawn still retain the full history of previous versions (see, e.g.,~\cite{ali2024qubo}). However, arXiv also provides the possibility to retrieve the submissions' source {\footnotesize \TeX{}} files. We show this in Figure~\ref{fig:arxiv_home}. In other words, \textit{anyone can download, and inspect, the files that a paper's author has uploaded to arXiv during the submission procedure}. We argue that this peculiarity deserves particular attention.

A {\footnotesize \TeX{}} project can be complex~\cite{tan2024inconsistencies}. For instance, besides the content shown in the PDF produced after the compilation, the source code of a {\footnotesize \TeX{}} project can include custom macros or even textual comments that are not part of the actual PDF. Moreover, focusing the attention on research papers, it is well-known that ``a scientific paper consists of a constellation of artifacts that extend beyond the document itself''~\cite{usenix2024artifact}. Therefore, it is sensible to assume that the {\footnotesize \TeX{}} source files of a research paper contain a variety of data (beyond textual comments) that are not necessary to produce the paper, such as unused images, old sections of the paper, or even files that are completely unrelated to the PDF (e.g., experimental source code, research spreadsheets, or different PDF documents). Indeed, nowadays, online {\footnotesize \TeX{}} editors such as Overleaf (which has over 20M users~\cite{overleaf2025about}) enable sharing projects across different users, facilitating joint writing between researchers~\cite{nacke2023write} who, during their interactions, may put such ``unnecessary'' data within the files of a {\footnotesize \TeX{}} project.

Simply put, a large variety of data not necessary to make the final PDF can be placed within the source {\footnotesize \TeX{}} files of a scientific article. As long as such data remains within the authors' control (e.g., on a shared Overleaf project), then no harm is done. Similarly, if the paper is to be uploaded on arXiv, the authors can diligently  clean their {\footnotesize \TeX{}} project from any data that is not needed to make the PDF or that is not intended to be publicly released. 
Yet, we wonder: what if authors of arXiv's submissions ``forget'' to clean their projects before uploading the source {\footnotesize \TeX{}} files on arXiv? And what if the authors are oblivious of the fact that the {\footnotesize \TeX{}} files of their arXiv submissions are  publicly retrievable? In either of these what-if scenarios, there is a risk that some arXiv submissions may contain \textit{residual data}, i.e., data that is not needed to produce the PDF. Such residual data not only {\small \textit{(i)}}~wastes storage space on arXiv's servers, but may also {\small \textit{(ii)}}~contain sensitive information that the authors do not want to be made public---such as undisclosed research details, private data, or embarrassing\,/\,harmful text.

\vspace{1mm}
\noindent
\textsc{\textbf{Research Goal and Major Findings.}} 
We hypothesize that arXiv's submitters may not clean their {\footnotesize \TeX{}} source files before upload. We thus seek to answer two research questions: ``\textit{how much residual data is there on arXiv?}'' (RQ1) and ``\textit{what sort of sensitive information can be found within such residual data?}'' (RQ2). To answer these RQs, we carry out the largest longitudinal analysis of arXiv submissions' source files. We download the source files of all \smamath{600}k submissions which appeared on arXiv within the first four months of the 2015--2025 timespan (i.e., 11 years). Then, we develop \texttt{BaRDE} (short for ``Bulk arXiv Residual Data Extractor''), a custom tool that we use to process all the source files we downloaded (which add up to 1.6TB in compressed format). 

Thanks to \texttt{BaRDE}, we can quantitatively answer RQ1: after decompressing our sample (obtaining 2.1TB of data), \textbf{we found 584GB of residual data, i.e., 27\% of the data uploaded (and stored) on arXiv is not necessary to create the corresponding paper}. Worryingly, we find that the situation worsened over the years: the percentage of residual data was \smamath{\approx}14\% in 2015--2017, and rose to over 30\% after 2020 (with a peak at 32\% in 2022). Moreover, the residual data for 4k {\footnotesize \TeX} projects represents more than 95\% of their total size. Altogether, such residual data puts a heavy strain on arXiv's storage. We have hence disclosed our findings to arXiv's leadership team, who confirmed receiving our message.

Then, to answer RQ2, we qualitatively analyzed a humanly-feasible portion of the residual data we found. Two researchers developed a codebook and analyzed the residual data of 200 randomly-chosen {\footnotesize \TeX} projects of 2025. We also carried out a keyword-driven search, looking for occurrences of specific terms across the textual comments of {\footnotesize \TeX} files. \textbf{Among the most concerning findings, we mention:} instances of offensive language (e.g., ``{\small \textsf{stupid fucking revision}}''), occasionally towards other authors (e.g., ``{\small \textsf{we should cite the stupid \anonym{} paper}}''); concealed text reporting links to code repositories not included in the actual paper; over 1.5k links to Google docs -- at least 200 of which ``accessible to anyone with the link'' and containing undisclosed/confidential data -- suggestions to avoid mentioning limitations; presence of private documents (e.g., theses under embargo, cover letters). 

Our findings suggest that the submitters of arXiv papers are not aware that the {\footnotesize \TeX} source files are publicly available. Hence, we have reached out to the authors of the submissions with concerning residual data that we found, informing them that: {\small \textit{(a)}}~the {\footnotesize \TeX} source files of their submissions are public and contain sensitive data (according to our judgment); {\small \textit{(b)}}~updating their submission by uploading new {\footnotesize \TeX} source files would overwrite those currently available on arXiv. Hence, with this paper, we aim to \textbf{raise awareness} and make members of the community aware of the public-availability of arXiv submissions' {\footnotesize \TeX} source files.

\vspace{0.5mm}
\noindent
\textsc{\textbf{Contributions.}} After downloading 1.6TB-worth of data~(§\ref{sec:preliminaries}), representing \smamath{\approx}600k submissions appeared on arXiv within the first four months of each year across 2015--2025:
\begin{itemize}[leftmargin=*]
    \item we develop \texttt{BaRDE} (§\ref{sec:method}), an original tool that enables bulk-extraction of \textit{residual data} (i.e., not necessary to produce the PDF of a {\footnotesize \TeX{}} project) of arXiv submissions;
    \item using \texttt{BaRDE}, we \textit{quantitatively analyze the residual data present on arXiv} (§\ref{sec:residual}), and we also validate our results (§\ref{sec:validation});
    \item through qualitative coding and keyword-driven searches, we manually analyze the residual data present on arXiv, focusing on cases raising \textit{privacy\,/\,confidentiality} concerns~(§\ref{sec:problematic}).
\end{itemize}
We responsibly disclosed our findings to arXiv and to authors of submissions with sensitive residual data (§\ref{sec:mitigation}). To the best of our knowledge, no prior work has carried out an analysis of similar size and scope on arXiv (related work is in §\ref{sec:related}).
\section{Preliminaries and Data Collection}
\label{sec:preliminaries}
\noindent
We summarize the submission process to arXiv (§\ref{ssec:submitting}), describe how we collected the data used for our research (§\ref{ssec:collection}), and explain how arXiv submissions are organized~(§\ref{ssec:organization}).

\subsection{Submitting Papers to arXiv}
\label{ssec:submitting}
\noindent
arXiv is a community-driven platform~\cite{arxiv2024report}. To make a submission on arXiv, users must first receive an endorsement (typically from another arXiv member~\cite{arxiv_endorsement}). Such an endorsement is, however, only valid for a specific \textit{category}. Indeed, arXiv hosts submissions pertaining to a variety of ``main'' scientific categories, such as Computer Science, or Physics; each of these categories is further broken down into ``specific'' categories, such as Cryptography \& Security, or Artificial Intelligence (see~\cite{arxiv_categories} for the complete categories).

arXiv's vision is to facilitate dissemination of ``established and emerging research''~\cite{arxiv2024report}. The submission of a paper is not bound to a peer-review process. Some submissions may present errors (and can be withdrawn~\cite{chawla2025withdrarxiv}) or can be just short drafts\,/\,critiques (e.g.~\cite{carlini2024cutting}); while others can become seminal works (e.g., the adversarial example paper by Goodfellow et al.~\cite{goodfellow2014explaining}). Some authors may upload on arXiv an early-version of a paper undergoing peer-review to share their discoveries as soon as possible~\cite{bagchi2025effects}, while others may use arXiv to provide a free version of a (peer-reviewed) article published in some journal or conference proceedings~\cite{moed2007effect}. 

To make a submission, it is necessary to upload its source files. As of November 2025, arXiv accepts two main classes of source files: a (zipped) file containing a \TeX{} project, which will be compiled on arXiv's infrastructure and used to produce the PDF hosted on arXiv; or a single PDF, representing the actual paper. While the latter may seem the most intuitive way, arXiv discourages such a practice, and explicitly forbids uploading PDF files generated via \LaTeX{}~\cite{arxiv_guidelines}. Nonetheless, as we wrote, arXiv enables anyone to download the source files of any given submission (see Figure~\ref{fig:arxiv_home}).

A submission's PDF remains on arXiv perpetually~\cite{arxiv_versioning}. Authors can update the submission by uploading different source files. Doing so will create a new ``version'' of a submission, resulting in a new PDF: downstream users can retrieve the PDF of any previous version of a submission. However, such versioning system does not apply for a submission's source files: whenever a submission is updated, its new source files overwrite the previous ones. Hence, users cannot retrieve the source files of past versions of a submission. 

\subsection{Retrieval of Papers (Data Collection)}
\label{ssec:collection}

\noindent
To comprehensively investigate what lies within the source files of arXiv submissions, we need to download thousands of submissions' source files. Doing so manually is unthinkable.

The best option we found\footnote{We first considered scraping. However, even by considering the ``scrape-friendly'' endpoint of arXiv (i.e.,~\cite{export_arxiv}), scraping approaches are not viable. According to arXiv's terms~\cite{arxiv_bulkdata} ``a reasonable rate are bursts of 4 requests per second with a 1 second sleep''; moreover, the robots page of arXiv explicitly prohibits programmatic source-file download~\cite{arxiv_robots}.}, which is also \textit{endorsed by arXiv itself}~\cite{arxiv_bulkdata}, was by accessing the copy of arXiv's database hosted on Amazon S3~\cite{arxiv_s3}. Such a database contains the source files of all arXiv's submissions, updated monthly, which we could download (at a cost~\cite{amazon_s3pricing}) from Amazon S3---without burdening arXiv's servers. Hence, after setting up an AWS account, we began (in May 2025) downloading the data used in our assessment. However, at this point in time, we did not know how much data we needed to download, nor how expensive such operations would be. Indeed, the only information provided in~\cite{arxiv_s3} was ``The source files [are grouped in] tar files of \smamath{\approx}500MB each and the complete set of source files is about 2.9 TB (March 2023)''. We did not know how many ``chunks'' of 500MB were included in each month, so it was impossible to make any estimate. 

Hence, we initially downloaded just the data for the first four months of 2025 (i.e., all months preceding the current one). This required 373GB (and around 40USD). We estimated that the size for the previous years would be inferior (see, e.g., the growth of submissions on arXiv~\cite{arxivstats}). To enable a fair analysis which could reflect yearly trends, and given that we had already downloaded the data for January--April 2025, we decided to focus only on the first four months of each year. According to our budget (around 150USD), we were able to afford downloads back to 2015. At the end of this process (which we finalized in July 2025), we downloaded a total of \smamath{\approx}1.6TB of data, distributed across 3,264 tar files of \smamath{\approx}500MB each. The detailed breakdown is in Table~\ref{tab:download}.

\begin{table}[t]
    \centering
    \caption{\textbf{Data downloaded from AWS S3.}
    \textmd{\footnotesize 
    Number of ``chunks'' and overall size (in GB) retrieved from S3 and containing the data we analyzed}} 
    \label{tab:download}
    \vspace{-4mm}
    \resizebox{0.9\columnwidth}{!}{
        \begin{tabular}{c|r|r|r|r|r|r|r|r||r|r}
            \toprule
            \textbf{Year} & \multicolumn{2}{c}{\textbf{January}} & \multicolumn{2}{c}{\textbf{February}} & \multicolumn{2}{c}{\textbf{March}} & \multicolumn{2}{c}{\textbf{April}} & \multicolumn{2}{c}{\textbf{Total}} \\
            & Chnks & Size & Chnks & Size & Chnks & Size & Chnks & Size & Chnks & Size \\
            \midrule
            2025 & 154 & 76.6 & 180 & 89.4 & 232 & 115.3 & 186 & 92.4 & 752 & 373.8 \\
            2024 & 134 & 67.3 & 149 & 74.9 & 176 & 88.0 & 161 & 81.0 & 620 & 311.3 \\
            2023 & 98 & 49.0 & 110 & 54.8 & 146 & 73.8 & 115 & 58.5 & 469 & 236.2 \\
            2022 & 84 & 42.5 & 90 & 45.0 & 122 & 61.7 & 98 & 49.1 & 394 & 198.4 \\
            2021 & 69 & 35.3 & 76 & 38.7 & 109 & 56.0 & 93 & 47.5 & 347 & 177.6 \\
            2020 & 40 & 21.1 & 42 & 23.1 & 50 & 27.3 & 51 & 27.8 & 183 & 99.4 \\
            2019 & 33 & 16.8 & 33 & 16.7 & 38 & 18.9 & 42 & 21.4 & 146 & 73.9 \\
            2018 & 27 & 13.2 & 28 & 14.1 & 32 & 16.1 & 32 & 15.8 & 119 & 59.3 \\
            2017 & 21 & 10.1 & 20 & 10.0 & 26 & 13.2 & 23 & 11.1 & 90 & 44.5 \\
            2016 & 17 & 8.5 & 19 & 9.3 & 21 & 10.6 & 21 & 10.2 & 78 & 38.7 \\
            2015 & 15 & 7.4 & 16 & 7.8 & 18 & 8.8 & 17 & 8.2 & 66 & 32.1\\ \midrule
            (sum) & 692 & 347.8 & 763 & 383.8 & 970 & 489.7 & 839 & 423.0 & 3,264 & 1,645.2 \\

            \bottomrule
        \end{tabular}
    }
    \vspace{-1mm}
\end{table}

\subsection{Structure of Submissions' Source Files}
\label{ssec:organization}
\noindent
After downloading our dataset, we inspected its contents to get a preliminary understanding of how to plan our analyses. 

\begin{figure}[t]
    \centering
    \includegraphics[width=0.9\columnwidth]{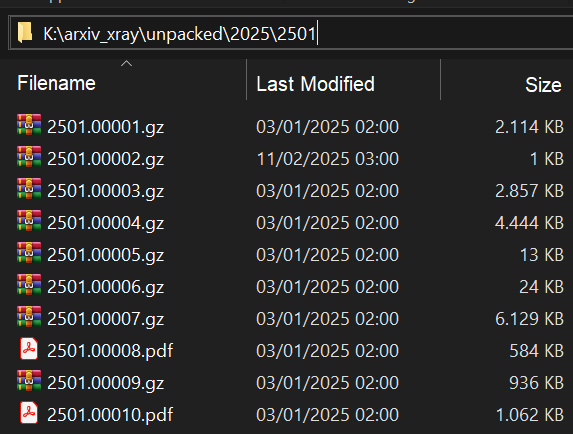}
    \vspace{-4mm}
    \caption{\textbf{Snippet of files extracted from a ``chunk''.}
    \textmd{\footnotesize We show the first 10 files (alongside their size and last modification), each denoting a specific submission, of the first ``chunk'' of January 2025 taken from S3.}} 
    \label{fig:subs}
    \vspace{-2mm}
\end{figure}

We extracted the 154 chunks of January 2025. We obtained a total of 19,407 files in the form ``YYMM.XXXXX.EXT'', where: YY and MM represent the year and month; XXXXX is a five-digit integer which progressively increases; and EXT is the extension of the file. Notably, the extension included only two types: PDF or GZ. We provide a snippet in Figure~\ref{fig:subs}. 

We hypothesized that all PDF files were submissions uploaded directly as a PDF: we confirmed such an hypothesis by visiting the web pages of such submissions (e.g., 2501.00008 in Figure~\ref{fig:subs}) and noticing that they lacked the ``Source Files'' button. We then turned our attention to the GZ files, i.e., compressed archives. Some were very small (e.g., 2501.00002 in Figure~\ref{fig:subs}) and contained a single extensionless file typically named ``withdrawn'': we hypothesized that such files were those referring to papers withdrawn from arXiv (we verified this---explaining why 2501.00002 has a more recent ``Last Modified'' date than the other submissions in Figure~\ref{fig:subs}). 

However, the vast majority of GZ files contained {\footnotesize \TeX{}}-related data, which could come in two different formats. Specifically, either {\small \textit{(a)}}~as a single {\footnotesize \TeX{}} file (e.g., this was the case for the content of 2501.00005.gz); or {\small \textit{(b)}}~as a blob which, if unpacked, yielded a ``full-fledged'' {\footnotesize \TeX{}} project, with various files and folders (e.g., this was the case for 2501.00001.gz). In either case, all GZ files always contained only one file (for the blob, such file was always named after the submission~ID).

\begin{figure*}[t]
    \centering
    \includegraphics[width=0.9\linewidth]{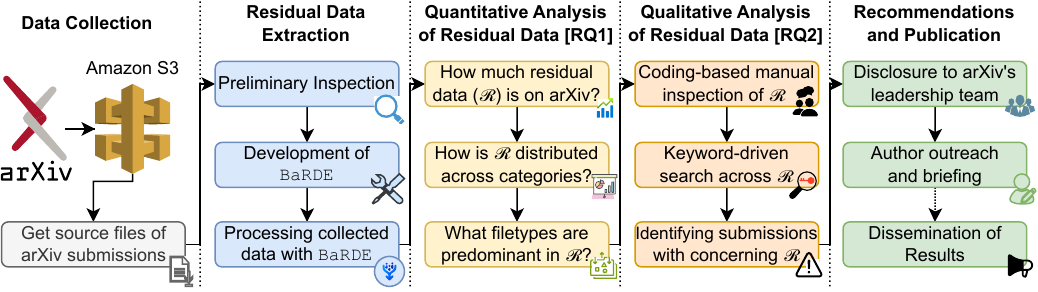}
    \vspace{-3mm}
    \caption{\textbf{Overview of our research.} \textmd{We downloaded our dataset in May--July 2025. We reached out to arXiv and to authors of ``problematic'' submissions in August 2025. We will wait at least 90 days (in line with best practices~\cite{projectzero}) before disseminating our findings.}}
    \label{fig:workflow}
    \vspace{-3mm}
\end{figure*}

\section{Research Methods and Tools}
\label{sec:method}
\noindent
We define the scope of our study and the challenges we need to overcome (§\ref{ssec:problem}). We then describe our solution, \texttt{BaRDE}~(§\ref{ssec:barde}), and finally run \texttt{BaRDE} on our dataset (§\ref{ssec:running}). 

The entire workflow of our study is shown in Figure~\ref{fig:workflow}.

\begin{cooltextbox}
\textsc{\textbf{Definitions.}} To facilitate the understanding of the remainder of our work, we define the most recurrent terms:
\begin{itemize}[leftmargin=*,noitemsep,nolistsep]
    \item \textit{Submission}: any ``paper\,/\,preprint'' for which an entry exists on arXiv, identified with an ID (e.g., 2501.00001).
    \item \textit{Project}: a submission whose source files are a {\footnotesize \TeX{}} project. 
    \item \textit{Residual data}: any data within a project that is not needed to produce the PDF on arXiv. This includes both files (e.g., unused images) and text (e.g., comments).
    \item \textit{Problematic projects}: projects containing residual data and which, in such data, include ``sensitive'' elements that can be considered as {\small \textit{(a)}}~offensive, {\small \textit{(b)}}~embarrassing, or which {\small \textit{(c)}}~leak private\,/\,confidential information.
\end{itemize}    
\end{cooltextbox}

\subsection{Problem Description and Challenges}
\label{ssec:problem}
\noindent
Our two RQs entail analyzing, quantitatively and qualitatively, the \textit{residual data} of an arXiv's submission source files. Of course, submissions whose source files are just a single PDF do not, by definition, have any residual data. Similarly, ``withdrawn'' submissions are not of interest for our study. 

\textbf{Tackling RQ1.} Our first objective is separating submissions whose source files resemble {\footnotesize \TeX{}} projects from those that are not. Then, we must identify the data that, in a project, is not necessary to produce the final PDF. Doing so, however, requires determining the \textit{root} {\footnotesize \TeX{}} file of a given project. Indeed, we do not know a priori what file is the ``starting point'' for the {\footnotesize \TeX{}} compilation. Nevertheless, given the scale of our dataset, the analysis of residual data must be done automatically (we cannot manually process 600k submissions!).

\textbf{Tackling RQ2.} After extracting the residual data of a given {\footnotesize \TeX{}} project, we must inspect such data to determine whether it is ``problematic'' or not. Such an objective can be done via some automation (e.g., keyword-driven searches or even via NLP techniques such as topic modeling~\cite{rogers2021primer}). Given the exploratory nature of our study (for which we lack ground truth) we shall not rely on black-box methods (due to lack of transparency~\cite{zini2022explainability} and/or ``hallucinations''~\cite{sriramanan2024llm}); instead, we prefer to use human expertise to infer whether the residual data of any given submission is truly worthy of attention. Such an approach implicitly introduces the reliance on best-effort strategies and prevents complete coverage.

\subsection{\texttt{BaRDE} {\normalsize (Bulk arXiv Residual Data Extractor)}}
\label{ssec:barde}
\noindent
We are not aware of any automated tool, whether open or closed-source, which enables us to carry out the analysis required to answer RQ1. A close match is the ``arxiv-latex-cleaner''~\cite{arxiv_cleaner} (ALC), but it cannot be used for our purpose because it assumes complete knowledge of a {\footnotesize \TeX{}} project---which we do not have. So, we had to develop our own tool, \texttt{BaRDE} (we will compare \texttt{BaRDE} with ALC in §\ref{sec:related}). 

We provide an overview of \texttt{BaRDE} (§\ref{sssec:design}) before presenting our design choices (§\ref{sssec:heuristics}). Finally, we discuss some technical difficulties encountered while developing \texttt{BaRDE}~(§\ref{sssec:technical}).

\subsubsection{Generic Design}
\label{sssec:design}

\noindent
\texttt{BaRDE} is the product of extensive trial-and-error done by the authors, who attempted to reverse-engineer not only the way in which arXiv organizes the source files of its submissions, but also the specific characteristics of arXiv itself (e.g., the ``ancillary folder''~\cite{arxiv_ancillary}). 

\textbf{Handling compressed data.} To develop \texttt{BaRDE}, we had to first consider that our dataset encompassed 1.6TB of data, the majority of which in \textit{compressed}  format (i.e., the GZ files). Moreover, inside each of these GZ files, the majority of {\footnotesize \TeX{}} projects were in the form of a blob---which must be unpacked before processing it. We could handle this ``two-step unpacking'' in two ways: {\small \textit{(a)}}~unpacking everything beforehand and then working on the unpacked data; or {\small \textit{(b)}}~keeping everything in compressed format, and deal with the unzipping\,/\,unpacking during runtime. We opted for the latter: the former would have required an unpredictable amount of storage space (at a minimum, assuming no compression, another 1.6TB) which we did not have. Such a design choice adds complexity, but it enables a smoother application of \texttt{BaRDE} by future research since it is designed to work on the source files in their natural format. To handle the unzipping\,/\,unpacking, we relied on 7zip~\cite{7zip}, which is free and supports both Windows and Linux OSes. 

\textbf{Focus on precision.} The other dilemma we faced when developing \texttt{BaRDE} was whether to opt for {\small \textit{(a)}}~coverage or {\small \textit{(b)}}~precision. We could either develop a tool that, in cases of uncertainty in determining the root file of a {\footnotesize \TeX{}} project, may have made a decision that could have raised ``false positives'' (e.g., flagging residual data that is actually an integral part of the {\footnotesize \TeX{}} project); or favor a more conservative approach. We favored precision: if \texttt{BaRDE} cannot determine the root file of a \TeX{} project, the submission is skipped (but a log is updated). Importantly, \texttt{BaRDE} operates statically and does not carry out {\footnotesize \TeX{}} compilation at runtime, which is a computationally demanding task (we validate this assertion in §\ref{sec:validation}).

\textbf{Workflow.} We provide the generic pseudocode of \texttt{BaRDE}'s workflow in Alg.~\ref{alg:barde} (extended by the functions in Alg.~\ref{alg:ext} in the Appendix). \texttt{BaRDE} inspects all files in a given \smamath{input\_folder} and stores the results of its analysis in dedicated \smamath{lists}: one for PDF files, one for \TeX{} projects, and another for files excluded from the analysis. If a file is a PDF, \texttt{BaRDE} updates \smamath{PDF\_list} and begins to inspect the next submission. If a file is a GZ file, \texttt{BaRDE} unpacks it and analyzes its contents---which are always a single file. Three cases can happen: 
\begin{itemize}[leftmargin=0.45cm]
    \item[{\small \textit{a)}}] the file is a blob, i.e., a {\footnotesize \TeX{}} project, for which it is necessary to infer its {\footnotesize \TeX{}} root and then infer its residual data (which can span both residual files and textual comments);
    \item[{\small \textit{b)}}] the file is a valid {\footnotesize \TeX{}} file (and, hence, the submission is a project), implying that the only residual data of this project are the potential comments included in this file;
    \item[{\small \textit{c)}}] the file is of an unrecognized type, or \texttt{BaRDE} cannot infer the {\footnotesize \TeX{}} root of a blob: the submission will be skipped.
\end{itemize}
While analyzing a (valid) {\footnotesize \TeX{}} project, \texttt{BaRDE} stores the following information: the corresponding $submission.ID$: the filenames and sizes of all of its $used$ and of its $residual$ files; and all the comments found in each $used$ {\footnotesize \TeX{}} file.

\SetKwRepeat{Do}{do}{while}%
\SetKw{KwBy}{by}%
\SetKw{KwOf}{in}%
\SetKw{KwIf}{if}%
\SetKw{KwIs}{is}%
\SetKw{KwAnd}{and}%
\SetKw{KwContinue}{continue}%
\SetKw{KwBreak}{break}
\begin{algorithm2e}[h]
    \caption{{\small \texttt{BaRDE} pseudocode (extended in Algorithm~\ref{alg:ext})}}
    \footnotesize
    \label{alg:barde}
    \DontPrintSemicolon
    \SetAlgoNoEnd

    {\setstretch{0.9}
    \KwIn{$input\_folder$ containing the submissions' source files (GZ and PDF) extracted from the chunks downloaded from S3.}}
    {\setstretch{0.9}
    \KwOut{$report$ containing, for each submission that is a valid {\tiny \TeX{}} project: statistics on its used \& residual files, as well as the concatenation of the textual comments found in its (used) {\tiny \TeX{}} files; and the submissions that have been excluded.}}
    
    \hrule
    
    $PDF\_list, TeX\_list, excluded\_list \gets$ emptyList();\\
    \For{submission $\KwOf$ $input\_folder$}{
        \If{submission.type $=$ "PDF"}{
            $PDF\_list \gets$ $submission.ID$;\\
            $\KwContinue$
        }
        {\tt \scriptsize // If here, then it's a GZ file, which must be unzipped}\\
        $content \gets$ unzip($submission$);\\

        {\tt \scriptsize // $content$ always has only one file}\\

            \If{content.type $=$ "TeX"}{
                $TeX\_list \gets$ singleTeX($content, submission.ID$);\\
                $\KwContinue$
            }
            \If{content.name $=$ submission.ID}{
                {\tt \scriptsize // $file$ is blob, which must be unpacked}\\
                $root, unpacked \gets$ inferRoot($content$);\\
                \If{root $= \emptyset$}{
                    $excluded\_list \gets$ $submission.ID$;\\
                    $\KwContinue$
            }
                $TeX\_list \gets$ multiTeX($root, unpacked, submission.ID$);\\
                $\KwContinue$
            }
            {\tt \scriptsize // Unrecognized file or withdrawn submission, skip}\\
            $excluded\_list \gets$ $submission.ID$;\\
        
        }
    $report \gets TeX\_list, PDF\_list, excluded\_list$;\\
    \Return $report$
\end{algorithm2e}

\subsubsection{Low-Level Heuristics for Analyzing \TeX{} Projects}
\label{sssec:heuristics}

\noindent
We explain the ideas behind three functions of \texttt{BaRDE}, shown in Alg.~\ref{alg:ext} (in the Appendix~\ref{app:barde}), which also provides additional low-level details on \texttt{BaRDE}: extractComments(\smamath{file}), inferRoot(\smamath{blob}), findResidual(\smamath{root, unpacked}).

\textbf{Extracting comments.} To determine text classifiable as ``comment'' in a {\footnotesize \TeX{}} file, we inspected the content of such a file and identified either: lines starting with \texttt{\%}; or text contained within {\small \texttt{\textbackslash begin\{comment\} \textbackslash end\{comment\}}}, {\small \texttt{iffalse~fi}}, or {\small \texttt{if0~fi}}, all of which being well-known methods to mark text that will not be shown in the final PDF (used also in~\cite{arxiv_cleaner}). By design, \texttt{BaRDE} extracts all comments of each \textit{used} {\footnotesize \TeX{}} file and stores them in a single file to ease further analyses.

\textbf{Finding the {\footnotesize \TeX{}} root.} Determining the root is not trivial. In theory, the ``starting point'' of a {\footnotesize \TeX{}} project is the file having, in the preamble, {\small \texttt{\textbackslash documentclass}}~\cite{latex_structure}. However, we found that many projects have \textit{multiple} {\footnotesize \TeX{}} files containing such a string in the preamble (this is typical when authors forget to remove {\footnotesize \TeX{}} files of ``templates''). Hence, merely looking for a file in the blob having {\small \texttt{\textbackslash documentclass}} is not enough to guarantee a root file. We thus implemented two heuristics to handle cases of projects having multiple files with {\small \texttt{\textbackslash documentclass}} in the preamble. First, we look for all files having {\small \texttt{\textbackslash documentclass}}, which we consider as \smamath{rootCandidates}. If, among these candidates, there is one file having the string ``main'' or ``paper'' or ``cameraready'' in its filename, then \texttt{BaRDE} considers this file as root; alternatively, if there is only one file contained in the topmost folder of the project, then such a file is the root. Otherwise, \texttt{BaRDE} skips the project, because it could not determine with certainty its root file.

\textbf{Identifying residual files.} After finding the root file (derived from a blob), \texttt{BaRDE} adopts a bottom-up approach to infer the project's residual files. First, \texttt{BaRDE} inspects the content of the root file, looking for occurrences of terms that denote methods ``calling'' a file, such as {\small \texttt{\textbackslash input}}, {\small \texttt{\textbackslash includegraphics}}, or {\small \texttt{\textbackslash include}} (we provide in the Appendix~\ref{sapp:patterns} the list of methods considered by \texttt{BaRDE}). Then, \texttt{BaRDE} captures the filename mentioned in such methods and stores it in a list called \smamath{seen}; afterwards, \texttt{BaRDE} will inspect such a file and recursively repeat the same process of finding ``called'' files, potentially updating the \smamath{seen} list. Finally, \texttt{BaRDE} will compare the filenames in \smamath{seen} against all the filenames included in the project: any filename in the project not included in \smamath{seen} is considered a residual file. However, to account for arXiv's ancillary files (which are files contained in a specific {\small \texttt{anc/}} folder created ad-hoc by the authors to enable users to freely inspect additional material related to their submission but not included in the PDF~\cite{arxiv_ancillary}), we remove any file included in the {\small \texttt{anc/}} folder (if present) from the list of residual files. 

\subsubsection{Technical Difficulties and Lessons Learned}
\label{sssec:technical}

\noindent
Developing \texttt{BaRDE} required engineering effort driven by our own expertise: we had to operate blindly because we were not aware of what can be found in each submission's source files. 

In particular, while developing \texttt{BaRDE}, we {\small \textit{(i)}}~used \smamath{\approx}2,000 projects as a blueprint to guide the low-level implementation of \texttt{BaRDE}, and we also {\small \textit{(ii)}}~used \smamath{\approx}1,100 submissions to validate the correctness of \texttt{BaRDE} (we discuss our validation in §\ref{sec:validation}, and limitations in §\ref{ssec:threats}). Indeed, even though navigating through a {\footnotesize \TeX{}} project may appear simple, there are a myriad of ``exceptions'' (especially given the absolute size of our dataset) that may cause issues. For instance, we were not aware of the arXiv-exclusive ``ancillary folder'', and we noticed it because we found a project (i.e.,~\cite{chen2025three}) having over 100MB of residual data, most of which included in an {\small \texttt{anc/}} folder, which made us~suspicious. 

Even realizing that the {\footnotesize \TeX{}} root file cannot be trivially inferred (due to the potential presence of duplicates) came to our surprise. For instance, we found cases in which \texttt{BaRDE} failed due to a clearly incorrect naming (e.g., we found a {\footnotesize \TeX{}} project consisting of a single file with extension ``.pdflatex'': we are surprised that such a file was correctly processed by arXiv servers). Moreover, we acknowledge that \texttt{BaRDE} in that it cannot handle ``aliases'': for instance, \texttt{BaRDE} is unable to recognize custom macros that call an external resource. Covering all such exceptions is beyond the scope of our work---but, as we will show, \texttt{BaRDE} is quite robust. 

Finally, an intrinsic issue we found was determining what constitutes a residual file \textit{for arXiv}. For instance, some projects had ``temporary'' files that are not referenced anywhere and which are generated at compile time, which technically are not required to be uploaded (since arXiv does so automatically) and persistently stored (we conjecture arXiv deletes such files after making the PDF).
Moreover, we found projects with many font-related files (mentioned in \texttt{.map} files), and we were unsure if such files were needed or not for arXiv's compilers (we ultimately decided that such files are \textit{not} residual because they are called by some {\footnotesize \TeX{}} files). Nonetheless, all such {\footnotesize \TeX{}}-related files, whether they are residual or not, are unlikely to {\small \textit{(i)}}~contain sensitive data and do not {\small \textit{(ii)}}~use a lot of storage space. Hence, our following analyses are not impacted by our design choices.

\begin{table}[t]
    \centering
    \caption{\textbf{Distribution of submissions in our sample.}
    \textmd{\texttt{BaRDE} analyzes each submission to determine if it is a PDF-only submission, or a \TeX{} project; we also report withdrawn submissions, or those which were skipped due to not being fully recognized by \texttt{BaRDE}.}} 
    \label{tab:preliminary_small}
    \vspace{-4mm}
    \resizebox{\columnwidth}{!}{
        \begin{tabular}{c||r|r|r|r|r|r}
            \toprule
            \textbf{Year} & \textbf{Submissions} & \textbf{Valid \TeX{} Projects} & \textbf{PDF-only} & \textbf{Withdrawn} & \textbf{Unclear Root} & \textbf{Unclear Type} \\
            \midrule
            2025 & 86,976 & 78,045 & 7,241 & 207 & 1412 & 71 \\
            2024 & 77,659 & 69,064 & 6,849 & 224 & 1458 & 64 \\
            2023 & 61,992 & 54,884 & 5,454 & 258 & 1328 & 68 \\
            2022 & 59,040 & 51,942 & 5,468 & 214 & 1320 & 96 \\
            2021 & 58,817 & 51,418 & 6,028 & 200 & 1101 & 70 \\
            2020 & 54,382 & 47,736 & 5,421 & 205 & 940 & 80 \\
            2019 & 48,889 & 43,064 & 4,844 & 188 & 677 & 116 \\
            2017 & 38,133 & 33,991 & 3,518 & 187 & 325 & 112 \\
            2016 & 36,087 & 32,291 & 3,208 & 199 & 243 & 146 \\
            2015 & 33,524 & 29,751 & 3,213 & 190 & 257 & 113 \\ \midrule
            \textbf{(agg)} & 599,613 (100\%) & 531,203 (88.6\%) & 55,638 (9.3\%) & 2,231 (0.3\%) & 9,504 (1.6\%) & 1,037 (0.1\%) \\

            \bottomrule
        \end{tabular}
    }
    \vspace{-2mm}
\end{table}

\subsection{Running \texttt{BaRDE} on our Dataset}
\label{ssec:running}

\noindent
We ran \texttt{BaRDE} across our dataset of 1.6TB. We present the results in Table~\ref{tab:preliminary_small} (extended in Table~\ref{tab:preliminary} in the Appendix).

\textbf{Overview.} Overall, \texttt{BaRDE} found that, out of 599,613 total submissions, 55,638 (9.3\%) are PDF-only files---confirming that such a format is rarely used. \texttt{BaRDE} also found that 2,231 (0.3\%) submissions have been withdrawn (we counted the number of ``excluded projects'' whose filename was ``withdrawn''). Finally, \texttt{BaRDE} correctly processed a total of 531,203 {\footnotesize \TeX{}} projects---which will be the subject of our analyses.

\textbf{Performance assessment.} We can use these results to estimate the ``coverage'' of \texttt{BaRDE}. 
Barely 10k projects (out of 600k, i.e., 1.7\%) could not be processed by \texttt{BaRDE} due to either having an unclear root (9,504, 1.6\%) or raising other issues (e.g., projects using the deprecated {\small \texttt{\textbackslash documentstyle}}) which prevented analyses (1,037, 0.1\%). Hence, \texttt{BaRDE}'s effectiveness was not hindered by our choices (we further validate \texttt{BaRDE}'s performance in §\ref{sec:validation}).

\textbf{Operational runtime.} We measured the time required to process our dataset with \texttt{BaRDE}. Altogether, it took \smamath{\approx}42 hours to process all of our dataset, indicating that \texttt{BaRDE} could analyze an average of 3.5 submissions per second. These results have been measured on an AMD Ryzen 5800X3D (@4.5GHz), with 32GB of RAM. However, we make two observations. First, the majority of the runtime is due to I/O operations needed to unpack each compressed GZ-file\,/\,blob, given that it is a procedure that entails storing the extracted data in a temporary folder: these operations are very time consuming, and their scale was big enough that it could potentially damage an SSD drive~\cite{ssd_lifecycle}, which is why we did them on an HDD (with 7200rpm). Using a faster storage drive would substantially decrease the runtime. Second, we ran \texttt{BaRDE} by using a single thread of our CPU. However, \texttt{BaRDE} does not have concurrency requirements and can be freely run by launching it multiple times, each using a dedicated thread and processing a subset of our dataset, thereby significantly increasing the speedup. For instance, had we run \texttt{BaRDE} by leveraging all 16 threads of our CPU and by specifying a dedicated storage device (to avoid I/O bottlenecks) we would have processed our dataset in less than 4 hours. Hence, we argue that \texttt{BaRDE} is, computing wise, an efficient solution to carry out large-scale analyses of arXiv submissions' source files (a claim we factually support in §\ref{ssec:validation_remarks}).

\begin{table}[!t]
    \centering
    \caption{\textbf{Size (in MB) of Residual Data ($\mathcal{R}$).} \textmd{We report the cumulative size (computed across the first four months of each year) of the residual files, comments, as well as the total size of $\mathcal{R}$ , and total size of projects; and the ratio between the last two elements.}} 
    \label{tab:residual_main}
    \vspace{-3mm}
    \resizebox{\columnwidth}{!}{
        \begin{tabular}{c||r|r|r|r|r}
            \toprule
                
            \textbf{Year} & \begin{tabular}{c} Residual \\ Files size\end{tabular} & \begin{tabular}{c} Size of \\ Comm. \end{tabular} & \begin{tabular}{c} Total size \\ of $\mathcal{R}$ \end{tabular} & \begin{tabular}{c} Tot. size \\ Projects \end{tabular} & \% $\mathcal{R}$ \\ \midrule
            2025 & 125,815 & 746 & 126,561 & 465,648 & 27.18\% \\
            2024 & 114,641 & 667 & 115,308 & 411,000 & 28.06\% \\
            2023 & 95,123 & 504 & 95,627 & 305,768 & 31.27\% \\
            2022 & 82,328 & 469 & 82,797 & 257,847 & 32.11\% \\
            2021 & 75,031 & 444 & 75,475 & 241,110 & 31.30\% \\
            2020 & 31,792 & 385 & 32,177 & 129,361 & 24.87\% \\
            2019 & 17,656 & 320 & 17,976 & 95,650 & 18.79\% \\
            2018 & 12,828 & 267 & 13,095 & 80,907 & 16.19\% \\
            2017 & 8,949 & 221 & 9,170 & 63,737 & 14.39\% \\
            2016 & 8,831 & 200 & 9,031 & 61,530 & 14.68\% \\
            2015 & 7,367 & 168 & 7,535 & 54,589 & 13.80\% \\ \midrule
            
            (agg) & 580,365 & 4,391 & 584,756 & 2,167,152 & 26.98\% \\

            \bottomrule
        \end{tabular}
    }
    \vspace{-4mm}
\end{table}

\section{Residual Data on arXiv [RQ1]}
\label{sec:residual}
\noindent
We focus on our first research question: quantifying the residual data on arXiv. To this end, we analyze the reports generated by \texttt{BaRDE} for submissions that are valid {\footnotesize \TeX{}} projects. 

We first provide an overview~(§\ref{ssec:residual_overview}). Then, we break-down our results by considering the distribution of residual data across scientific categories~(§\ref{ssec:residual_categories}). Finally, we conclude by analyzing the file-types that compose residual data (§\ref{ssec:residual_types}).

\subsection{Overview of Residual Data}
\label{ssec:residual_overview}
\noindent
\textit{How much residual data is on arXiv?} To answer RQ1, we consider the most straightforward quantitative metric: the~size.

\textbf{Main results.} We report the results of our primary analysis in Table~\ref{tab:residual_main}. Specifically, for each year, we show: the total size of residual \textit{files}, the total size of textual \textit{comments} found in (used) {\footnotesize \TeX{}} files, the total size of residual data (given by summing the previous two elements), the total size of the projects, and the percentage of residual data w.r.t. the total size of a project. (The distribution across months is provided in Table~\ref{tab:residual_main-months} in the Appendix). Overall, there are 584GB of residual data, of which 4.3GB are comments, and 580GB are files not needed to make the final PDF. In contrast, the total size of projects is of 2.1TB (note: all of these numbers refer to the size of data \textit{uncompressed}). Therefore, in our sample, 26.98\% of the data is not required for the PDF compilation.

\begin{table}[t]
    \centering
    \caption{\textbf{Distribution of projects according to the total size of their residual files ($\mathcal{F}$).} \textmd{Note that $\mathcal{R}$=$\mathcal{F}$+Comments.}} 
    \label{tab:residual_distribution_bytes}
    \vspace{-3mm}
    \resizebox{0.9\columnwidth}{!}{
        \begin{tabular}{c||c|c|c}
            \toprule

            Year & $\mathcal{F}$<1KB & 1KB<$\mathcal{F}$<1MB & $\mathcal{F}$>1MB \\   
            
            \midrule

            2025 & 20,273 (25.98\%) & 40,564 (51.98\%) & 17,208 (22.05\%) \\
            2024 & 19,979 (28.93\%) & 34,410 (49.82\%) & 14,675 (21.25\%) \\
            2023 & 17,950 (32.71\%) & 23,685 (43.15\%) & 13,249 (24.14\%) \\
            2022 & 18,900 (36.39\%) & 20,871 (40.18\%) & 12,171 (23.43\%) \\
            2021 & 20,562 (39.99\%) & 19,765 (38.44\%) & 11,091 (21.57\%) \\
            2020 & 22,692 (47.54\%) & 17,614 (36.90\%) & 7,430 (15.56\%) \\
            2019 & 27,693 (64.29\%) & 10,766 (25.00\%) & 4,613 (10.71\%) \\
            2018 & 27,262 (69.87\%) & 8,346 (21.39\%) & 3,409 (8.74\%) \\
            2017 & 25,186 (74.10\%) & 6,357 (18.70\%) & 2,448 (7.20\%) \\
            2016 & 24,662 (76.37\%) & 5,704 (17.66\%) & 1,925 (5.96\%) \\
            2015 & 23,470 (78.89\%) & 4,767 (16.02\%) & 1,514 (5.09\%) \\ \midrule
            (agg) & 248,629 (46.80\%) & 192,849 (36.30\%) & 89,733 (16.89\%) \\

            \bottomrule
        \end{tabular}
    }
\end{table}

\begin{figure}[!tbp]
    \vspace{-3mm}
    \centering
    \includegraphics[width=\linewidth]{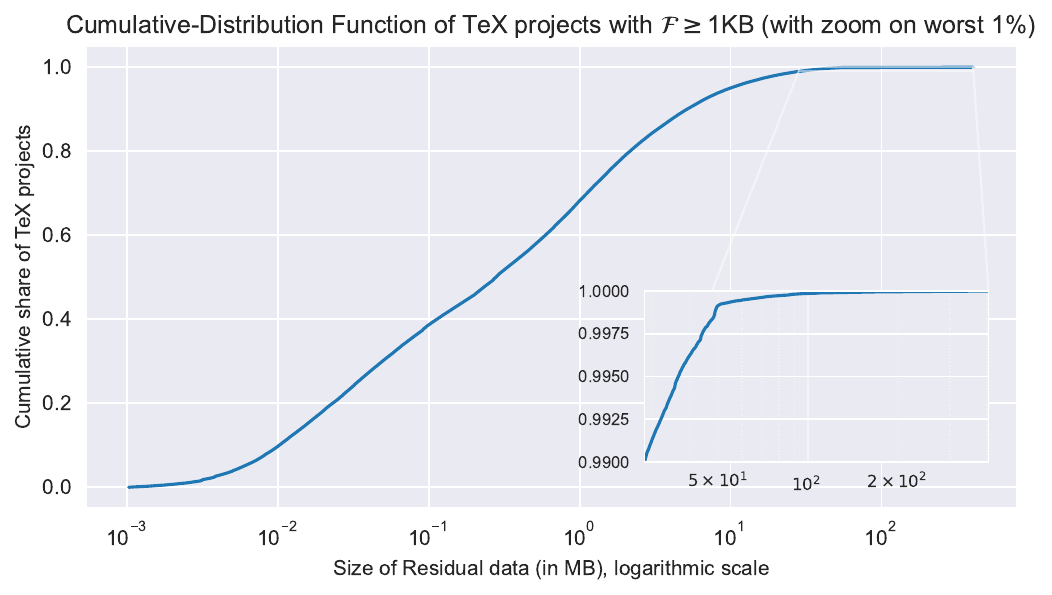}
    \vspace{-7mm}
    \caption{\textbf{Distribution of projects with size of $\mathcal{F}$$\geq$1KB (count: 285,582).} \textmd{Some projects (38) have more than 100MB worth of $\mathcal{F}$.}}
    \label{fig:cdf}
    \vspace{-3mm}
\end{figure}

\begin{figure*}[!htbp]
    \centering
    \begin{subfigure}[t]{0.49\textwidth}
        \centering
        \includegraphics[width=\columnwidth]{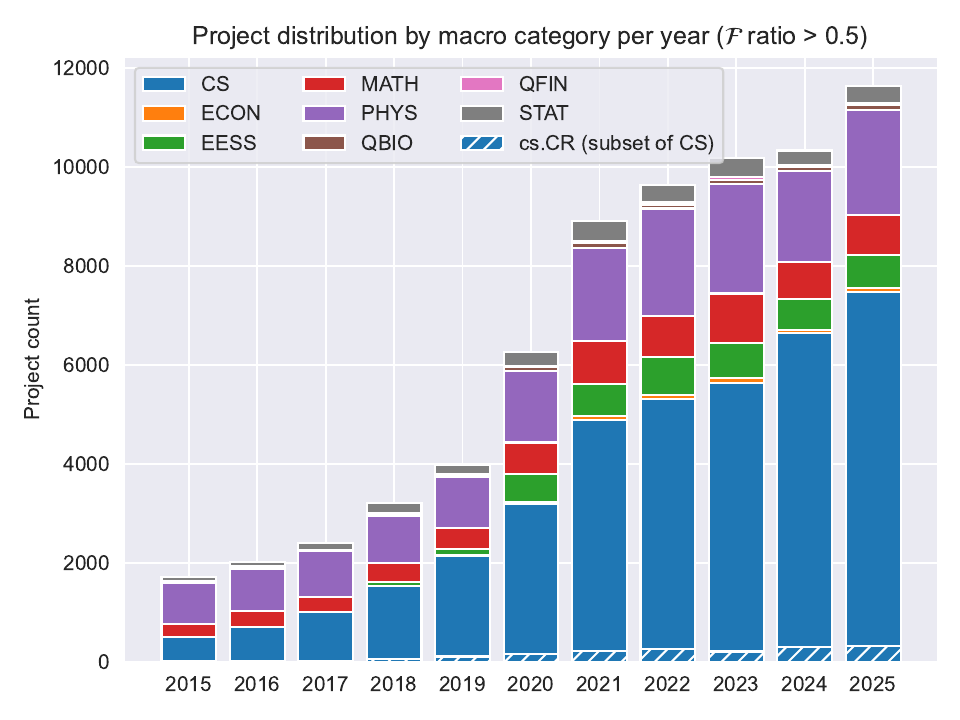}
    \end{subfigure}%
    ~ 
    \begin{subfigure}[t]{0.49\textwidth}
        \centering
        \includegraphics[width=\columnwidth]{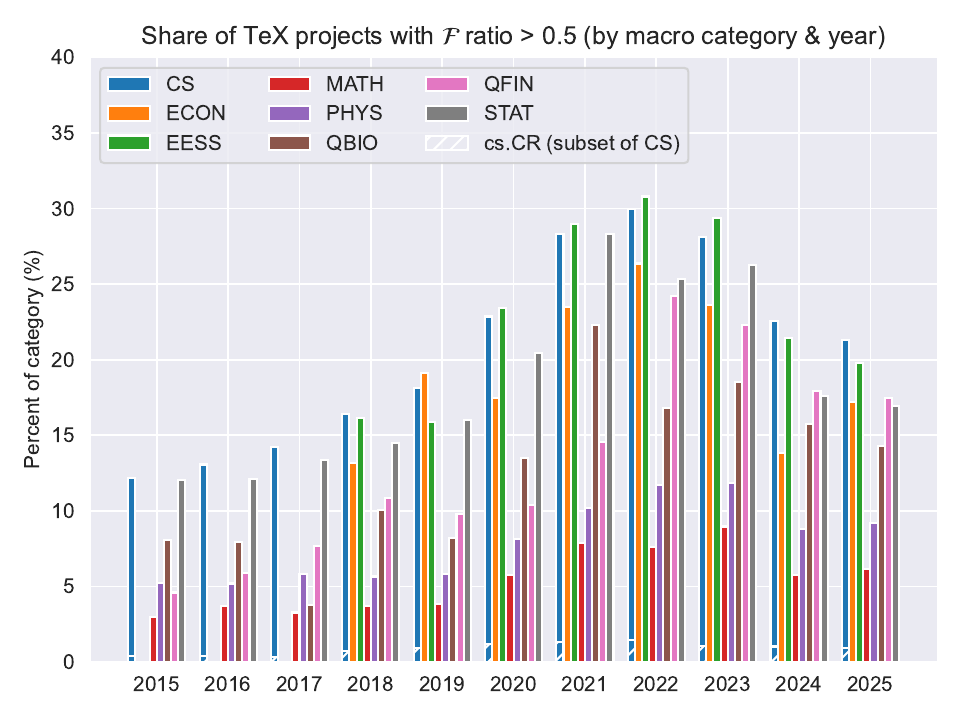}
    \end{subfigure}
    \vspace{-3mm}
    \caption{Distribution (absolute:left, relative:right) across scientific categories of projects whose size of $\mathcal{F}$ is >50\% of the project~size.}
    \label{fig:half_categories}
    \vspace{-3mm}
\end{figure*}

\textbf{Size-wise distribution of residual files.} Our previous results show that there are a lot of residual data in terms of overall size. However, how is such residual data distributed across projects? To investigate this, we considered, for each year, the \textit{number of projects} having: {\small \textit{(i)}}~less than 1KB-worth of residual files, {\small \textit{(ii)}}~between 1KB and 1MB of residual files, and {\small \textit{(iii)}}~above 1MB of residual files. The results are shown in Table~\ref{tab:residual_distribution_bytes} (for which we report in Table~\ref{tab:residual_distribution_bytes-months} the details across months). We see that even though the residual files for \smamath{\approx}248k (\smamath{\approx}47\%) projects add up to less than 1KB (which can be negligible), there are \smamath{\approx}90k (\smamath{\approx}17\%) projects that have more than 1MB worth of residual files. To better visualize these phenomena, we show in Figure~\ref{fig:cdf} the cumulative distribution function for submissions having more than 1KB worth of residual files, with a zoom on the ``worst'' 1\%. We found 38 projects that have more than 100MB worth of residual data.

\textbf{Relative size-wise distribution.} We complement our previous results by showing, in Table~\ref{tab:residual_distribution_relative} (extended by Table~\ref{tab:residual_distribution_relative-months} in the Appendix), the number of projects whose cumulative size of residual files represents [less than 5\%; between 5\% and 50\%; between 50\% and 95\%; above 95\%] of the total project size. We found 71,076 (13.38\%) projects whose residual files represent more than 50\% of their total size; in particular, for 4,031 (0.76\%) projects, 95\% of their size is made up by files not needed to produce the final PDF.

\begin{table}[!tbp]
    \centering
    \caption{\textbf{Distribution of projects according to: (total size of residual files)\,/\,(total project size).} \textmd{$\mathcal{F}$=``size of residual files''.}} 
    \label{tab:residual_distribution_relative}
    \vspace{-3mm}
    \resizebox{\columnwidth}{!}{
        \begin{tabular}{c||c|c|c|c}
            \toprule            
            Year & $\mathcal{F} <$ 5\% & 5\%$\leq \mathcal{F} <$50\% & 50\%$\leq \mathcal{F} <$95\% & $\mathcal{F}\geq$95\% \\  
            
            \midrule

            2025 & 41,337 (52.97\%) & 25,067 (32.12\%) & 11,088 (14.21\%) & 553 (0.71\%) \\
            2024 & 37,429 (54.19\%) & 21,300 (30.84\%) & 9,638 (13.96\%) & 697 (1.01\%) \\
            2023 & 29,476 (53.71\%) & 15,098 (27.51\%) & 9,825 (17.90\%) & 485 (0.88\%) \\
            2022 & 28,673 (55.20\%) & 13,497 (25.98\%) & 9,334 (17.97\%) & 438 (0.84\%) \\
            2021 & 29,706 (57.77\%) & 12,656 (24.61\%) & 8,611 (16.75\%) & 445 (0.87\%) \\
            2020 & 30,698 (64.31\%) & 10,689 (22.39\%) & 6,085 (12.75\%) & 264 (0.55\%) \\
            2019 & 31,687 (73.57\%) & 7,303 (16.96\%) & 3,864 (8.97\%) & 218 (0.51\%) \\
            2018 & 30,278 (77.60\%) & 5,475 (14.03\%) & 3,029 (7.76\%) & 235 (0.60\%) \\
            2017 & 27,578 (81.13\%) & 3,957 (11.64\%) & 2,227 (6.55\%) & 229 (0.67\%) \\
            2016 & 26,875 (83.23\%) & 3,353 (10.38\%) & 1,841 (5.70\%) & 222 (0.69\%) \\
            2015 & 25,366 (85.26\%) & 2,637 (8.86\%) & 1,503 (5.05\%) & 245 (0.82\%) \\ \midrule
            (agg) & 339,103 (63.84\%) & 121,032 (22.78\%) & 67,045 (12.62\%) & 4,031 (0.76\%) \\

            \bottomrule
        \end{tabular}
    }
    \vspace{-3mm}
\end{table}

\subsection{Residual Files Across Categories}
\label{ssec:residual_categories}

\noindent
Given that arXiv accepts submissions pertaining to various scientific categories, we wondered which category has submissions with the higher amount/percentage of residual files. 

\textbf{Method.} We downloaded the dataset (available at~\cite{arxiv_metadata}) containing the complete metadata of all arXiv submissions. Such details include, among others, the specific categories assigned to each submission. We hence cross-referenced the categories to the projects we found having residual data. For simplicity, we only considered the 8 main categories (CS=Computer Science, ECON=Economics, EESS=Elec. Engineering and System Sciences, MATH=Mathematics, PHYS=Physics, QBIO=Quant. Biology, STAT=Statistics, QFIN=Quant. Finance~\cite{arxiv_categories}), and we also highlight projects in ``Cryptography\&Security'' (CR, subcategory of CS). Given that we are interested in the ``higher amount\,/\,percentage'', we only consider submissions with more than 1MB worth of residual files or a ratio of residual files (w.r.t. total project size) above 0.5.

\textbf{Results.} We report the results in Fig.~\ref{fig:half_categories} (focusing on projects with over half of their size being of residual files) and Fig.~\ref{fig:onemb_categories} (in the Appendix, focusing on projects with >1MB worth of residual files). We can see (looking at the plot on the left in each figure) that, starting from 2019, \textit{CS is the category with the highest number of projects with more than half of their size being of residual files, or with more than 1MB of residual files} (the CR subcategory does not seem to be a heavy contributor in this regard). From a relative viewpoint (plot on the right), however, things are slightly different: \textit{CS is still tends to be the ``worst''} (e.g., starting from 2019, over 20\% of projects in CS have more than half of their size being of residual files), but there are other categories (most notably, EESS, which is thematically close to CS) with similar numbers.

\begin{figure}[!t]
    \centering
    \includegraphics[width=\columnwidth]{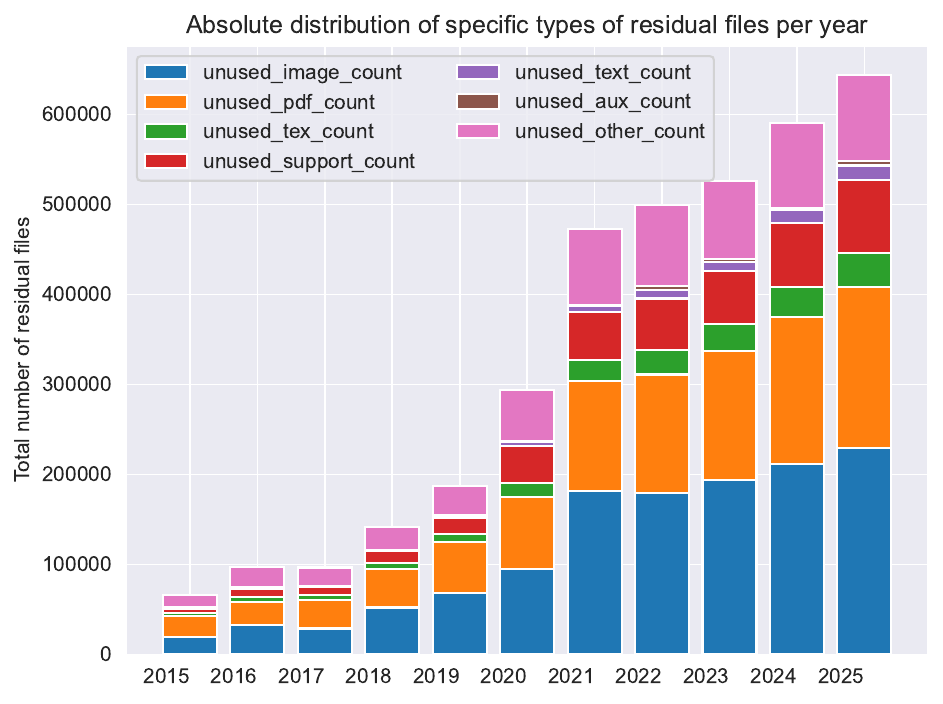}
    \vspace{-6mm}
    \caption{Absolute number of file types across the residual files.}
    \label{fig:types_count}
    \vspace{-3mm}
\end{figure}

\subsection{Types of Residual Files}
\label{ssec:residual_types}

\noindent
We conclude our quantitative analysis by inspecting the types of files that are superfluous for making the final PDF.

\textbf{Approach.} Given that \texttt{BaRDE} stores the list of all residual files for each project, we used the \textit{extension} of such files to create seven distinct groups: images (e.g., .svg, .png), PDF (.pdf), {\footnotesize \TeX{}} (.tex), support (e.g., .sty, .cls), text (e.g., .txt, .md), auxiliary (e.g., .enc, .tfm), and other (anything else); we report in Appendix~\ref{sapp:type} the full mapping. We then computed the yearly distribution of each of these groups in terms of total occurrences, and total size. The intention is identifying which file-types are mostly prevalent across residual files.

\textbf{Results.} We report the results in the plots shown in Figure~\ref{fig:types_count} and Figure~\ref{fig:types_size} (in the Appendix). From Figure~\ref{fig:types_count}, it is evident that the ``image'' and ``pdf'' categories are the most prevalent, followed by files in the ``support'' and ``other'' category. However, by looking at Figure~\ref{fig:types_size}, we see that, size-wise, files in the ``image'' category take significantly (verified with a t-test at \smamath{p<.05}) more storage space. For instance, just in 2025 there are nearly 70GB worth of ``images'', compared to \smamath{\approx}45GB of ``pdf'', \smamath{\approx}10GB of ``other'', and \smamath{\approx}2GB of ``support''. 

\begin{cooltextbox}
    \textsc{\textbf{Takeaways.}} Across our sample (first four months of 2015--2025), we found 584GB of residual data on arXiv, representing 27\% of the size of its {\footnotesize \TeX{}} projects. Computer Science is the category which contributes the most to these numbers. Images are the most prevalent types of residual files.
\end{cooltextbox}

\section{Validation of \texttt{BaRDE}}
\label{sec:validation}

\noindent
Recall that, when processing a {\footnotesize \TeX{}} project, \texttt{BaRDE} uses our custom heuristics (§\ref{sssec:heuristics}) to {\small \textit{(i)}}~determine the root {\footnotesize \TeX{}} file and {\small \textit{(ii)}}~infer residual files. We have already verified (§\ref{ssec:running}) that, across our dataset, \texttt{BaRDE} could not process \smamath{\approx}1.7\% of the {\footnotesize \TeX{}} projects due to an unclear root file. In this section, we attempt to quantify the cases in which our heuristics failed, leading to \texttt{BaRDE} making an incorrect decision. 

We do so via a validatory experiment carried out on a different dataset. We first discuss the data collection and generic approach (§\ref{ssec:validation_dataset}), then report the results (§\ref{ssec:validation_results}), and finally make some remarks (§\ref{ssec:validation_remarks}).

\subsection{Validation Dataset (and Approach)}
\label{ssec:validation_dataset}
\noindent
To fairly validate our results, we must use a different dataset. Moreover, to provide an accurate validation, we must manually check the ground truth of each sample in such a dataset---meaning that our validation cannot span a dataset with the same size (i.e., 600k submissions) as the one of our study.

To this end, we downloaded (from the AWS S3 bucket~\cite{arxiv_s3}) the first ten ``chunks'' of source files (\smamath{\approx}5GB compressed) referring to arXiv submissions of May 2025. Altogether, these chunks contained the source files of 1,104 submissions. We then extracted/unpacked all such source files (which required 4 minutes on our system). Out of these 1,104 submissions, 91 are PDF-only submissions, whose cumulative size is of 432MB); the remaining 1,013 submissions have a cumulative size of 5.53GB uncompressed, and of 4.59GB in GZ format.

To validate our root-file heuristic, we seek to {\small \textit{(i)}}~run \texttt{BaRDE} on this validation dataset and then use the ``root'' file inferred by \texttt{BaRDE} to {\small \textit{(ii)}}~compile the corresponding PDF, and finally {\small \textit{(iii)}}~compare such PDF with the PDF of the respective submission available on arXiv: cases in which the two PDFs differ are those in which our root-file selection heuristic failed. Therefore, we downloaded the PDF (also from the S3 bucket~\cite{arxiv_s3}) of each of these 1,104 submissions (totaling 4.41GB). We found that, for 3 submissions, there was no PDF: these were cases of ``withdrawn'' papers.

Afterwards, to validate the residual-files selection process, we will {\small \textit{(i)}}~remove the residual files flagged by \texttt{BaRDE} from the source files of the project, and {\small \textit{(ii)}}~attempt to compile the PDF: if the compilation fails, it means that \texttt{BaRDE} incorrectly flagged a file that was necessary for the compilation. 

\subsection{Validation Results}
\label{ssec:validation_results}
\noindent
We processed our validation dataset with \texttt{BaRDE}, which required 5 minutes. We confirmed that all 91 PDF-only files were correctly recognized as such. Moreover, \texttt{BaRDE} correctly identified the three ``withdrawn'' papers; and \texttt{BaRDE} skipped 2 projects (0.18\%) which did not have a valid {\footnotesize \TeX{}} structure (they only had a single .txt file), as well as 23 projects (2\%) which had multiple root candidates. These results align with those of our large-scale analysis (see Table~\ref{tab:preliminary_small}). 

We validated our root-file identification heuristic. We took the source files of the 985 projects for which \texttt{BaRDE} inferred the root {\footnotesize \TeX{}} file and compiled them on our systems. This process took 4 hours. Then, we checked the PDF against the one on arXiv. We found only one case in which \texttt{BaRDE} inferred the incorrect root {\footnotesize \TeX{}} file (the correct root was ``Draft.tex'', but \texttt{BaRDE} used ``main.tex''). In other words, our heuristic was correct in 984 out of 985 cases (i.e., 99.9\%).

We validated the residual-file selection heuristic. After removing all residual files flagged by \texttt{BaRDE} from each project, we attempted the PDF compilation. We found only 9 (out of 985, i.e., 0.9\%) cases in which the compilation failed due to missing files removed by \texttt{BaRDE}, all related to alias / custom commands (e.g., one project defined the macro {\small \texttt{\textbackslash image}} to replace {\small \texttt{\textbackslash includegraphics}}). This is a known limitation which we acknowledged (see §\ref{sssec:technical}). However, even in these cases, \texttt{BaRDE} still correctly identified other residual files.

\subsection{Observations and Remarks}
\label{ssec:validation_remarks}
\noindent
We use this validation experiment to shed light and factually justify the overall effectiveness of \texttt{BaRDE}.

First, the results of \texttt{BaRDE} on our validation dataset resemble that of our large-scale analysis, indicating that these two datasets have a similar distribution, enabling us to extend the results of our validation to our main analysis (in §\ref{sec:residual}).

\texttt{BaRDE} can make errors in 1\% of the analysed projects, leading to incorrectly flagging files as residual (i.e., ``false positives''). However, these cases are due to custom/special commands and/or projects that do not follow {\footnotesize \TeX{}} best practices. Still, even in these cases, \texttt{BaRDE} still correctly identified other residual files (e.g., only the files called with the custom command are affected). We hence conclude that such errors have a negligible impact on our quantitative results.

Finally, as we anticipated (in §\ref{ssec:running}) carrying out the aforementioned analysis is much more computationally expensive than running \texttt{BaRDE}. Recall that processing our main dataset (of \smamath{\approx}600k submissions) took \smamath{\approx}42 hours, whereas compiling the PDF for the validation dataset (of \smamath{\approx}1k submissions) took \smamath{\approx}4 hours. In other words, if we wanted to be certain of our results by compiling the PDF via {\footnotesize \TeX{}}, it would have required 2400 hours, i.e., 57\smamath{\times} the time required to use \texttt{BaRDE}. We therefore consider that an error rate of \smamath{\approx}1\% of our heuristics is well compensated by \texttt{BaRDE}'s computational efficiency.

\section{Problematic Projects on arXiv (RQ2)}
\label{sec:problematic}
\noindent
To identify ``problematic'' projects, we carried out manual analyses. We first inspected all residual data of a randomly-chosen subset of our sample (§\ref{ssec:random}), followed by a keyword-driven search on comments (§\ref{ssec:keyword-comments}) and on residual files (§\ref{ssec:file}).

{{\setstretch{0.95}
\small 
\begin{softtextbox}
    \textsc{\textbf{Disclaimer.}} What makes a project ``problematic'' is subjective. For instance, some people may not see any problem in the presence of a swear word in a comment; and some authors may have deliberately left some ``undisclosed research data'' in their submissions.\footnote{Note that many people regret sharing some information~\cite{wang2011regretted,rash2019disconnect}, since it may backfire years later~\cite{parker2019should}. We also consider these circumstances.} Our analyses are driven by our own judgment: \textit{we} would not want that our submissions included the data we found in any ``problematic project'' discussed in this section.
\end{softtextbox}
}}

\subsection{Random Sampling (Full Manual Check)}
\label{ssec:random}
\noindent
Analyzing all the residual data we found in our sample manually is clearly an unfeasible task. To get a preliminary understanding of what can be found within the residual data of arXiv submissions' source files, we performed a complete search across a small, randomly-chosen subset of our sample. 

\textbf{Approach.} We adopted the dual-reviewer system~\cite{stoll2019value}: two researchers (with >5 years of research experience) independently analyzed the same projects, with the goal of determining if such projects contained residual data that could be considered as ``problematic''. After the analysis, discussions were done to reach a consensus. The analysis would encompass all residual data (comments\,+\,residual files) for these submissions.
For a humanly-feasible analysis, we set our ``budget'' to 200 submissions; we opted for those in 2025 because they are the most recent ones and, in case we found some problems, authors are more likely to receive our emails (see §\ref{ssec:outreach}). Hence, for an unbiased selection, we randomly sampled 50 submissions for each month in 2025; we did not specify any criteria in terms of size\,/\,ratio of residual data. 

\textbf{Codebook.} To guide our analyses, we defined a codebook through which we could classify each submission according to the ``problem'' we found. The codebook was finalized after analyzing 100 submissions: we then uniformly reapplied our codes to these 100 submissions, and used the codes for the remaining 100. The codebook encompassed the following categories, for which we provide a brief explanation: 
\begin{itemize}[leftmargin=*]
    \item \textit{Author exchange.} Covers instances of comments that indicate conversations among authors, sometimes anticipated by a command denoting a specific person (e.g., ``{\small \texttt{\textbackslash AuthorX\{}\textsf{not sure what to put here}\texttt{\}}}'') or not (e.g., ``\textsf{\small Revise.}'');
    \item \textit{Direct translation.} Instances wherein comments include long sequences of text in a language different from English which, if translated via automatic tools (e.g.,~\cite{deepl}), yield an almost-perfect match with the text in the paper's PDF;
    \item \textit{Inappropriate language.} Covers instances wherein comments contain offensive terms (e.g., ``{\small \textsf{fuck}}'') or derogatory statements towards other entities (e.g., ``{\small \textsf{stupid \anonym{} paper}}'');
    \item \textit{Data leak.} Instances in which the project (either in the comments or in the residual files) contains research data not disclosed in the paper, personal (and not public) details about the authors, or clearly confidential information (e.g., documents not publicly available, such as cover letters);
    \item \textit{Other.} For all other cases (e.g., comments including ``acknowledgments'' not mentioned in the paper's PDF).
\end{itemize}
We believe the aforementioned codes are a clear indicator of ``problematic'' projects. If the reader is unsure why such codes could be considered as problematic, we discuss in the Appendix~\ref{sapp:codebook} how each code can lead to harmful consequences towards the authors of the corresponding submission.

\textbf{Problematic (verified) findings.} Our final results are as follows. Overall, we found problems in 40\,/\,200 (20\%) projects. Specifically, we coded: 16 (8\%) projects in the ``author exchange'' category, 6 (3\%) in ``direct translation'', 1 (0.5\%) in ``inappropriate language'', 6 (3\%) in ``data leak'', 11 (5.5\%) in ``other''. We provide some (anonymized) excerpts in the Appendix~\ref{sapp:excerpts_random}. For Excerpt~\ref{exc:four}, its paper never mentions release of code (either before or ``after publication''); whereas for Excerpt~\ref{exc:eleven}, there is no acknowledgment mentioned in the paper (and the acknowledged entities are not mentioned). Excerpts~\ref{exc:seven} and~\ref{exc:eight} (whose links are accessible) inspired our future analyses. We did not find any specific problem in the remaining 160 (80\%) projects. Note: even in these cases, there could still be data that the authors do not want to be public.

\subsection{Keyword-Driven Search (on Comments)}
\label{ssec:keyword-comments}
\noindent
We used our previous findings as a scaffold to carry out a broader analysis leveraging some automated mechanisms. We used a simple and reliable technique: keyword searches.\footnote{We tried using ``topic models'', such as ToxicBERT~\cite{ranasinghe2023text}, but they always yielded false positives (e.g., papers with a lot of ``\%'' would be deemed as highly offensive), so we decided not to rely on these black-box techniques.}

\textbf{Approach.} We defined a list of keywords and then automatically searched across all comments of each project (among the 531k of our dataset), flagging all those that had at least one match for each of the keywords we considered. The list was derived after internal discussion among the authors of this paper, based on their own experience with the English language. The list (with keywords) is as follows:
\begin{itemize}[leftmargin=*]
    \item \textit{Offensive}: fuck, shit, dumb, idiot, bastard, crap, stupid;
    \item \textit{Derogatory}: terrible, horrible, mess, garbage, trash, useless;
    \item \textit{TODO}: todo, fixme, tbd;
    \item \textit{Exclamation}: wtf, geez, lmao;
    \item \textit{URI}: docs.google.com/, github., gitlab., C:\textbackslash Users, /home/;
    \item \textit{Hidden prompts:} positive review only (from~\cite{nikkei2025positive}).
\end{itemize}
Importantly, we acknowledge that the mere presence of one of these keywords does not make the project problematic.\footnote{For instance, papers about ``offensive language'' (e.g.,~\cite{caselli2020feel}) can have valid reasons to have commented-out portions of the text mentioning ``fuck''.} This is why we carried out manual checks on a subset of the flagged papers to ensure that the ``context'' in which a keyword was mentioned truly indicates a problematic project. 

\textbf{High-level results.}
We report the results in the format [keyword; \# of projects mentioning it at least once]: [fuck; 108], [shit; 279], [dumb; 270], [idiot; 54], [bastard; 24], [crap; 141], [stupid; 554], [terrible; 418], [horrible; 235], [mess; 984], [garbage; 567], [trash; 241], [useless; 1,975], [todo; 41,122], [fixme; 1,747], [tbd; 2,187], [wtf; 131], [geez; 5], [lmao; 20], [docs.google.com/; 828], [github.; 22,075], [gitlab.; 201], [C:\textbackslash Users; 126], [/home/; 1715], [positive review only; 3]. After some manual checks, we realized the high number of ``todo'' was because many projects use the {\small \texttt{todonotes}} package; whereas the high occurrences for ``mess'' were due to some {\footnotesize \TeX{}} templates (e.g.,~\cite{aaai25_template}) mentioning it (and the corresponding portions were left as comments).

\textbf{Problematic (verified) findings.} Even though not all the projects mentioning any of such terms were problematic, we found many instances of clearly offensive or derogatory statements which -- we believe -- may lead to harm towards the authors if publicly shared. We report some (anonymized) excerpts in the Appendix~\ref{sapp:excerpts_keyword}. Notably, the presence of one of such keywords may indicate \textit{more problematic} comments (see, e.g., Excerpts~\ref{exc:4},~\ref{exc:5}, or~\ref{exc:10}). Nonetheless, we also decided to carry out extensive manual checks specifically focused on the ``docs.google.com'' keyword. First, we re-did our keyword search, this time counting all \textit{unique occurrences} of such term, which yielded 1,510 matches (spread across 828 distinct projects). Then, we focused on the 314 links pertaining to the 158 projects of 2025: we manually inspected all of these links to infer if such links pointed to documents ``accessible to anyone with the link''. We found that 200\,/\,314 (64\%) of the links are freely accessible (as of August 2025), and they clearly contained confidential data (see also §\ref{ssec:outreach}). 

\subsection{Inspection of Residual Files}
\label{ssec:file}
\noindent
Lastly, we try to pinpoint problematic projects by looking at the unused files, and then inspecting their contents. Similarly to our previous analysis, we do this via a keyword-search.

\textbf{Approach and High-level Results.}
We craft a script that looks at the various ``reports'' produced by \texttt{BaRDE}, looking for mentions of the specific terms across the residual files. The terms and their total occurrences across our sample are:
\begin{itemize}[leftmargin=*]
    \item \!\textit{Code}:\,{\small [.exe;\,35],\,[.sh;\,1,884],\,[.py;\,5,259],\,[.bat;\,230],\,[.ipynb;\,828]}
    \item \!\textit{Documents}: {\small [.doc; 76], [.docx; 1,086], [.xlsx; 1,447], [.xls; 133], [.ppt; 61], [.pptx; 240]}
    \item \!\textit{Video}: {\small [.mp4; 843], [.avi; 81], [.mov; 31]}
    \item \!\textit{Misc}: {\small[cover\_letter; 314], [rebuttal; 5,964], [reviews; 567]}.
\end{itemize}
After manual inspection of some matches, we confirmed these were all ``residual files'' (validating \texttt{BaRDE}'s output). 

\textbf{Problematic (verified) findings.} We found projects having cover letters (including a recommendation letter, see Excerpt~\ref{exc:13}), rebuttals (some including comments, see Excerpt~\ref{exc:14}), or previous reviews---all of which being confidential data that is not meant to be available to the entire world. Some instances of ``documents'' included preliminary drafts or experimental details not mentioned in the paper. As a byproduct of these (and other file-related) inspections, we also found a project with a thesis still under embargo (only the abstract is available online). 

\begin{cooltextbox}
    \textsc{\textbf{Takeaways.}} Our manual inspection on a randomly chosen subset of 200 projects revealed that 1\,/\,5 contained problematic residual data. Keyword-driven searches with string ``docs.google.com'' across comments yielded 1,510 hits: we manually checked the 314 occurring in 2025, and found that 64\% of these URLs point to freely accessible documents containing confidential data. Manual investigations focused on residual files yielded concerning findings, such as projects containing private documents and letters.
\end{cooltextbox}

\section{Threats to Validity and Discussion}
\label{ssec:threats}
\noindent
Our conclusions depend on the ability of \texttt{BaRDE} to extract residual data (files\,+\,comments) in arXiv submissions.

We have discussed (in §\ref{ssec:running}) that \texttt{BaRDE} is not able to analyze submissions that, despite being {\footnotesize \TeX{}} projects, do not have a clear root file; or which have a structure that does not align with modern {\footnotesize \TeX{}} versions (e.g., incorrect extensions or deprecated methods). In these cases, which we quantified as representing 1.7\% of our dataset, \texttt{BaRDE} would not proceed with its analysis to avoid raising false positives.

We have further validated (in §\ref{sec:validation}) the accuracy of our heuristics: accordingly, \texttt{BaRDE} incorrectly flags some residual files for 1\% of the analysed {\footnotesize \TeX{}} projects: most of these cases are due to alias/special commands\footnote{Note that such special cases would also affect tools such as ALC~\cite{arxiv_cleaner}.} that can hardly be parsed with a static tool such as \texttt{BaRDE}, since it operates on compressed files and does not carry out {\footnotesize \TeX{}} compilation: this is not a limitation, but rather a strength, since compiling {\footnotesize \TeX{}} projects is resource intensive (we estimated running \texttt{BaRDE} takes $\frac{1}{57}$ of the time otherwise required to compile the PDF).

To sum up, we believe that potential misattributions by \texttt{BaRDE} to be in a low-enough number to not threaten our quantitative conclusions (in §\ref{sec:residual}), albeit we acknowledge the actual numbers of residual data in our sample may be slightly lower. The conclusions of our manual analyses (in §\ref{sec:problematic}) are correct: whenever we encountered a ``problematic'' project, we manually checked the source files of such project to verify that it truly contained our identified problem.

\section{Mitigation and Reflections}
\label{sec:mitigation}
\noindent
We have brought to light (in §\ref{sec:problematic}) a subtle problem, for which we cannot quantify its complete prevalence. In an attempt to mitigate this problem, we discuss here some real-world considerations---starting from our outreach campaign.

\subsection{Outreach}
\label{ssec:outreach}
\noindent
Our findings pertain to two groups of stakeholders: arXiv as a platform, because residual data harms their servers; and the authors of submissions with problematic residual data.

We \textbf{reached out to arXiv} in early August 2025. We sent the email reported in Email~\ref{emailbox:arxiv} (in the Appendix) to the arXiv leadership team~\cite{arxiv_leadership}. We could not find the addresses of two members, but one member forwarded our mail to the last two members, thereby confirming having received our email. 

Then, starting from the second-half of August 2025, we \textbf{reached out to the authors} of ``problematic'' submissions. Overall, we sent 161 emails, covering {\small \textit{(a)}}~the projects analyzed in our random sampling; {\small \textit{(b)}}~projects with freely-accessible links to Google documents; and {\small \textit{(c)}}~all projects for which we included an excerpt in this paper. We sent the email reported in Email~\ref{emailbox:authors} (in the Appendix). 
Reaching out to the authors was not simple: it cannot be done automatically (arXiv prohibits programmatic access to the show-email endpoint~\cite{arxiv_robots}) so we had to retrieve the email addresses manually. For submissions having more more than one address on the paper's front page, we reached out to multiple authors (after consultation with our ethical board); otherwise, we only sent the email to the address of the submitter shown on arXiv.

We received 35 replies (a response rate of 22\%). 
Among these, only two respondents confirmed that they wanted to make the submission exactly as it was available on arXiv: in contrast, and worryingly, 27 respondents admitted that it was not their intention to disclose all source files, and took action. 
As of October 2025, 36 submissions for which we sent an email had their source files updated \textit{after our disclosure}.

Finally, we sent a follow-up to the arXiv leadership team in October 2025, stating that some authors may overlook that the arXiv source files are public. We suggested recommendations (e.g., those discussed in §\ref{ssec:recommendations}) and expressed our willingness to help.

\subsection{Real-World Implications}
\label{ssec:implications}
\noindent
We examine the issue we uncovered under a real-world lens.

\textbf{Retroactivity.} arXiv has recently updated its submission-upload system which now explicitly tells authors that some files may not be needed~\cite{arxiv_legacy}. This may help reducing the size of residual files, but: {\small \textit{(a)}}~it does not help for removing ``comments'' in future submissions; and, crucially, {\small \textit{(b)}}~it does not help for existing submissions. Indeed, the issue we brought to light affects all submissions on arXiv, since everyone can download their source files---which may not be ``clean''. Therefore, arXiv's new submission-upload system does not represent a universal solution to the public-availability of arXiv submissions' source files.

\textbf{Public availability.} Our qualitative findings suggest that arXiv submitters may not be aware that the source files of their submissions are publicly available. Indeed, even though it is undeniable that residual data may put a burden on arXiv (and hence it is reasonable that arXiv tries to limit this by, e.g., suggesting which files may not be needed), our research underscored that the public-availability of arXiv submissions' source files poses a privacy risk---due to lack of awareness. We point out that, in arXiv's FAQs~\cite{arxiv_whytex}, it is stated ``What if my TeX source has potentially embarrassing self-comments in it? Well... you should probably take them out.'' Hence, \textit{arXiv does warn users} of their submissions' source files being public; however, it is evident that some users did not read the warning, or consider\,/\,understand its broader implications.

\textbf{Other Platforms.} arXiv is the largest eprint-focused platform for which the source files of its submissions are publicly available. Platforms such as bioRxiv~\cite{biorxiv}, medRxiv~\cite{medrxiv}, or techRxiv~\cite{techrxiv} do not enable such a possibility. Users of such platform are therefore not exposed to the privacy risks underscored in this work.

\subsection{Recommendations}
\label{ssec:recommendations}
\noindent
To improve the status quo, we have three recommendations for arXiv, and one for the entire arXiv's community.

First, arXiv should not make a submission' source files publicly available \textit{by default}. We believe that the privacy risks of the indiscriminate public-availability of source-files outweigh its benefits---especially given the potential lack of awareness on this matter. We assert arXiv should implement an ``opt-in'' mechanism through which authors can express their willingness to share their source files. 

Second, arXiv should \textit{reach out to the authors of all submissions currently on arXiv}. The submitters should be reminded that the source files are publicly available, and that they can update their submissions to delete content that is not meant to be public (and of the existence of the {\small \texttt{anc/}} folder~\cite{arxiv_ancillary}).

Third, arXiv should better emphasize -- during the submission process (and not just in the FAQs) -- that the source files of a submission will be publicly available. Potentially, the authors should fill a textual box with ``I am aware that all source files of my submission will be publicly available'' (which is more effective than, e.g., ticking a checkbox after scrolling through some ToS). Such a warning/advisory can be complemented by integrating some of \texttt{BaRDE}'s functions in the submission upload system, such as the comment-extraction pipeline. Such an integration will enable the authors to quickly observe the ``residual data'' of their submission, including comments, and take an informed decision of whether to keep such data or remove (parts of) it. 

Finally, the authors of arXiv's submissions should broadcast our message to their colleagues. The sooner a submission is updated, the less likely its authors are going to suffer damaging consequences due to potentially-sensitive data included in their submissions' source files.

\section{Related Work}
\label{sec:related}

\noindent
We are not aware of any work with a similar scope as ours. 

First, and as a disclaimer, for our paper we did not rely on an ``exposed AWS S3 bucket'' (see, e.g.,~\cite{el2025file,continella2018there}). It is arXiv itself that \textit{points researchers to the specific AWS S3 bucket} to retrieve its submissions in bulk~\cite{arxiv_s3}, which can be freely used for research purposes~\cite{arxiv_tou}, which is our case.

The closest works entail papers seeking to, e.g., analyze comments in code repositories (e.g., GitHub~\cite{farber2020analyzing,jarczyk2014github,chatziasimidis2015data,alfieri2024exploring}) or identifying privacy leaks in code repositories (e.g.,~\cite{wen2022secrethunter,meli2019bad,feng2022automated}); but no work has carried out a systematic analysis of arXiv submissions with a specific focus on their source files.

Some works discuss the pros and cons of using arXiv~\cite{rastogi2022arxiv} (e.g., whether it improves citation metrics~\cite{shuai2012scientific,bagchi2025effects}) or analyzed arXiv submissions from a graph-analytics perspective~\cite{clement2019use}, but these are all goals orthogonal to ours.

Even in terms of scale, our study is larger than those carried out by prior work that focused on arXiv (but with different goals). For instance, Lin et al.~\cite{lin2020many} analyzed how many computer-science preprints submitted within 2008--2017 have been printed, and their study encompassed only 142k submissions (i.e., their dataset is 1\,/\,4 of ours); whereas Movva et al.~\cite{movva2024topics} analyze trends across 17k submissions focusing on LLMs (i.e., their dataset is 1\,/\,35 of ours). 

From a technical viewpoint, our tool also provides novel functionalities w.r.t. the closest open-source tool we could find, ALC~\cite{arxiv_cleaner}, which we discuss Appendix~\ref{ssec:comparison}.

\section{Conclusions}
\label{sec:conclusions}
\noindent
We analysed 1.6TB of data, encompassing 600k submissions appeared on arXiv in the first four months of 2015--2025. 

We found that 27\% of the data of {\footnotesize \TeX{}} projects is not needed to produce the PDF. Qualitative analyses on a subset of projects with residual data revealed privacy-noteworthy findings, such as undisclosed research details, or the presence of derogatory statements. Altogether these results indicate that arXiv submitters may not be aware that the source files of arXiv submissions can be downloaded by anyone.

We reached out to arXiv, as well as to the authors of some submissions for which we found privacy-noteworthy residual data. However, the problem we brought to light remains open, and only a policy change by arXiv (e.g., removing the ability to download submissions' source files) or a community-driven awareness effort (to notify authors that their submissions' source files are public but can be updated and overwritten) would mitigate this problem.

\textbox{\textbf{arXiv's response.} We did not receive any reply to our email in Oct. 2025, so we sent a third follow-up on Dec. 31st 2025, in which we attached a draft of this paper (thereby showing our complete procedure and findings), and explictly asked if there was any intention in reaching out to authors. We received a response on Jan. 10th 2026, which appreciated our efforts, but noted that there was no intention to reach out to individual authors en masse.}

\section*{Ethics}

\noindent
Our paper tackles various ethical themes. Let us discuss the ethics of our research. We anticipate that \textit{the authors consulted with the ethical office of their institutions about the ethics of this research, and it was deliberated that no ethical concern can arise from the methodology adopted in our study}. 

\subsection*{Ethics of our Methodology}
\noindent

First, our research is \textit{based on publicly-available data}. The authors of all arXiv's submissions have agreed to arXiv's terms of service before uploading their source files. Therefore, no rule\,/\,term\,/\,agreement\,/\,law is breached by analyzing, and using, such publicly-available data.

Second, the data collected for our research \textit{was obtained ethically}. We refrained from scraping arXiv, and---as recommended by arXiv itself~\cite{arxiv_s3}---opted for downloading our datasets from Amazon S3 (which led to costs we had to pay on our own). Hence, we did not violate arXiv's Terms~\cite{arxiv_tou}.

Finally, when reporting the outcome of our outreach campaign, we only provided aggregated data and it is impossible to identify any individual. Recipients to our email, who were made aware we were carrying out some research, have been treated ethically~\cite{bailey2012menlo}, and we acted in their own best interests. We refrained from wasting their time with unnecessary follow-ups unless explicitly told or unless it was apparent that they may have overlooked something (e.g., for freely-accessible Google documents, for which the authors thanked us after making them aware of this).

A remark can be made regarding our decision to send the email to multiple authors (if we found the email addresses). Let us explain this delicate issue. The process of submitting a paper to arXiv is overseen by only one author---who, in theory, must have permission by all co-authors. We envision a scenario wherein {\small \textit{(i)}}~an author submits a paper with ``problematic'' residual data inside, {\small \textit{(ii)}}~multiple authors receive our email, and {\small \textit{(iii)}}~the other authors ``complain'' against the submitting author for not having cleaned the source files before submission. We are aware of such a possibility, but we believe that the likelihood of this happening (given that we sent an email to multiple authors only for 1/3 of the cases) is low. Moreover, our intention when sending the email to multiple authors was to cover the case in which an email address may have been disabled, or the author left academia---in which case, it is important to warn also the other authors, who can take action if deemed necessary. Regardless, we emphasize that our methodology was chosen after a joint consultation with our institutions' ethical boards.

\subsection*{Ethics of our Publication}
\noindent
The publication of this paper would undeniably lead to our findings be announced to the entire World. The consequences of this can vary, depending on the outcome of decisions we cannot predict. In what follows, we discuss these cases and explain why the publication of this paper is the best course of action given our findings.

In an hypothetical scenario in which, possibly due to our message to the arXiv's leadership team, arXiv as a platform does not allow third-parties to download (either on arXiv, or on AWS S3) the submissions' source files, no harm will ensue by the publication of our work. Even if the world becomes aware that there is ``residual data on arXiv'', the fact that potential evildoers may want to repeat our analyses to cause harm to authors of arXiv's submissions would be prevented by lack of methods to obtain the source files. 

Alternatively, if arXiv does not change its policies w.r.t. the availability of submissions' source files, potential evildoers may be able to repeat our analyses. However, to minimize the potential harm, we have adopted the following precautions.
\begin{itemize}[leftmargin=*]
    \item First, we warned the authors of (some) submissions with problematic data. Such authors have hence the opportunity to update their submissions, preventing any harm;
    \item Second, we warned arXiv's team (who acknowledged receiving our email). This can lead to arXiv sending emails ``en-masse'' to inform authors to update their source files.
    \item Third, we will make \texttt{BaRDE} available \textit{only upon request}. Indeed, \texttt{BaRDE}'s capabilities can be used by attackers, too. However, by itself, \texttt{BaRDE} cannot cause any damage, since it requires access to the dataset of arXiv submissions' source files downloaded from AWS S3. If arXiv takes down such a dataset, we will open-source \texttt{BaRDE}. Otherwise, we will only provide it under explicit request by people with a reputable background. 
\end{itemize}
Regardless, and as an additional precautions to prevent harm, any excerpt included in this paper is provided in a format that makes it impossible to identify its authors (in particular, we do not reference the corresponding submissions), and we never explicitly pointed to any specific submission as having potentially-sensitive data.

\subsection*{Our Ethical Message}
\noindent
The issue we underscored is present \textit{now}. Attackers can already: ``steal'' confidential data carelessly included in the arXiv submissions' source files by some authors; or ``threaten'' to publicly-shame some authors due to derogatory\,/\,offensive\,/
embarrassing\,/\,controversial statements found in the respective comments. We believe the potential targets of such nefarious acts should be responsibly warned~\cite{kohno2023ethical}. 

Finally, we observe that publication of this work will lead to \textit{future authors of arXiv submissions} being aware of this problem. Hence, while we acknowledge that we---as researchers---do not have the means to fully solve this problem for submissions already present on arXiv, publication of this paper would prevent harm for future arXiv's submissions---which are increasingly growing in number.

\bibliographystyle{ACM-Reference-Format}

{

}

\appendices
\section{Extended Tables (and Extra Figures)}
\label{app:tables}
\noindent
We report the extension of the tables shown in the paper. Specifically, we report: 
Table~\ref{tab:preliminary} (extension of Table~\ref{tab:preliminary_small});
Table~\ref{tab:residual_main-months} (extension of Table~\ref{tab:residual_main});
Table~\ref{tab:residual_distribution_bytes-months} (extension of Table~\ref{tab:residual_distribution_bytes}); 
Table~\ref{tab:residual_distribution_relative-months} (extension of Table~\ref{tab:residual_distribution_relative}).

\begin{table*}[!htbp]
    \centering
    \caption{\textbf{Preliminary Results of our ``submission-type'' script.}
    \textmd{\footnotesize 
    After extracting all ``chunks'', \texttt{BaRDE} analyses the source files trying to determine if they denote PDF-only submissions, or \TeX{} projects; we also account for withdrawn or submissions which skipped due to not being fully recognized by \texttt{BaRDE}.}} 
    \label{tab:preliminary}
    \vspace{-2mm}
    \resizebox{2.1\columnwidth}{!}{
        \begin{tabular}{c||c|c|c|c|c?c|c|c|c|c?c|c|c|c|c?c|c|c|c|c?c|c|c|c|c?c|c|c|c|c}
            \toprule

            \multirow{2}{*}{\textbf{Year}}
            
            & \multicolumn{5}{c?}{\textbf{Submissions}} 
            & \multicolumn{5}{c?}{\textbf{Valid \TeX{} Projects}} 
            & \multicolumn{5}{c?}{\textbf{PDF-only}} 
            & \multicolumn{5}{c?}{\textbf{Withdrawn (certain)}} 
            & \multicolumn{5}{c?}{\textbf{Unclear Root}} 
            & \multicolumn{5}{c}{\textbf{Unclear Type}} 
            \\ \cmidrule{2-31}
            & Jan & Feb & Mar & Apr & TOT 
            & Jan & Feb & Mar & Apr & TOT 
            & Jan & Feb & Mar & Apr & TOT 
            & Jan & Feb & Mar & Apr & TOT 
            & Jan & Feb & Mar & Apr & TOT 
            & Jan & Feb & Mar & Apr & TOT \\
            \midrule
            2025 & 19407 & 21321 & 24391 & 21857 & 86976 & 17176 & 19196 & 21937 & 19736 & 78045 & 1842 & 1699 & 1968 & 1732 & 7241 & 49 & 72 & 49 & 37 & 207 & 330 & 338 & 413 & 331 & 1412 & 10 & 16 & 24 & 21 & 71 \\
            2024 & 18085 & 19481 & 20332 & 19761 & 77659 & 15832 & 17333 & 18184 & 17715 & 69064 & 1819 & 1736 & 1679 & 1615 & 6849 & 56 & 56 & 69 & 43 & 224 & 362 & 346 & 382 & 368 & 1458 & 16 & 10 & 18 & 20 & 64 \\
            2023 & 13870 & 14863 & 18249 & 15010 & 61992 & 12115 & 13248 & 16265 & 13256 & 54884 & 1365 & 1247 & 1483 & 1359 & 5454 & 61 & 65 & 67 & 65 & 258 & 311 & 286 & 411 & 320 & 1328 & 18 & 17 & 23 & 10 & 68 \\
            2022 & 13452 & 14038 & 17276 & 14274 & 59040 & 11669 & 12354 & 15308 & 12611 & 51942 & 1437 & 1263 & 1490 & 1278 & 5468 & 45 & 57 & 65 & 47 & 214 & 285 & 332 & 378 & 325 & 1320 & 16 & 32 & 35 & 13 & 96 \\
            2021 & 12747 & 13656 & 17273 & 15141 & 58817 & 11039 & 11965 & 15111 & 13303 & 51418 & 1422 & 1370 & 1748 & 1488 & 6028 & 49 & 50 & 62 & 39 & 200 & 221 & 256 & 328 & 296 & 1101 & 16 & 15 & 24 & 15 & 70 \\
            2020 & 12010 & 12932 & 14415 & 15025 & 54382 & 10456 & 11441 & 12658 & 13181 & 47736 & 1293 & 1189 & 1439 & 1500 & 5421 & 42 & 50 & 60 & 53 & 205 & 202 & 232 & 243 & 263 & 940 & 17 & 20 & 15 & 28 & 80 \\
            2019 & 11537 & 11302 & 12654 & 13396 & 48889 & 10171 & 9962 & 11108 & 11823 & 43064 & 1130 & 1106 & 1299 & 1309 & 4844 & 45 & 46 & 56 & 41 & 188 & 165 & 157 & 161 & 194 & 677 & 26 & 31 & 30 & 29 & 116 \\
            2018 & 10609 & 10593 & 11560 & 11352 & 44114 & 9382 & 9397 & 10132 & 10106 & 39017 & 1064 & 1039 & 1229 & 1062 & 4394 & 38 & 43 & 44 & 34 & 159 & 100 & 92 & 126 & 125 & 443 & 25 & 22 & 29 & 25 & 101 \\
            2017 & 9186 & 8910 & 11008 & 9029 & 38133 & 8117 & 7970 & 9880 & 8024 & 33991 & 931 & 814 & 936 & 837 & 3518 & 44 & 40 & 56 & 47 & 187 & 66 & 67 & 104 & 88 & 325 & 28 & 19 & 32 & 33 & 112 \\
            2016 & 8251 & 9142 & 9746 & 8948 & 36087 & 7368 & 8170 & 8757 & 7996 & 32291 & 736 & 830 & 804 & 838 & 3208 & 46 & 43 & 74 & 36 & 199 & 63 & 62 & 59 & 59 & 243 & 38 & 37 & 52 & 19 & 146 \\
            2015 & 7912 & 8054 & 9191 & 8367 & 33524 & 7013 & 7129 & 8118 & 7491 & 29751 & 776 & 794 & 924 & 719 & 3213 & 45 & 47 & 49 & 49 & 190 & 56 & 62 & 72 & 67 & 257 & 22 & 22 & 28 & 41 & 113 \\
            \midrule
            (overall) & \multicolumn{5}{c?}{599613 (100\%)} & \multicolumn{5}{c?}{531203 (88.59\%)} & \multicolumn{5}{c?}{55638 (9.28\%)} & \multicolumn{5}{c?}{2231 (0.37\%)} & \multicolumn{5}{c?}{9504 (1.59\%)} & \multicolumn{5}{c}{1037 (0.17\%)}\\

            \bottomrule
        \end{tabular}
    }
\end{table*}

\begin{table*}[!htbp]
    \centering
    \caption{\textbf{Size (in MB) of Residual Data ($\mathcal{R}$).} \textmd{We report the cumulative size (computed across the first four months of each year) of the residual files, comments, as well as the total size of $\mathcal{R}$ , and total size of projects; and the ratio between the last two elements.}} 
    \label{tab:residual_main-months}
    \vspace{-3mm}
    \resizebox{2.1\columnwidth}{!}{
        \begin{tabular}{c||c|c|c|c|c?c|c|c|c|c?c|c|c|c|c?c|c|c|c|c?c|c|c|c|c}
            \toprule

            \multirow{2}{*}{\textbf{Year}}
            
            & \multicolumn{5}{c?}{\textbf{January}} 
            & \multicolumn{5}{c?}{\textbf{February}} 
            & \multicolumn{5}{c?}{\textbf{March}} 
            & \multicolumn{5}{c?}{\textbf{April}} 
            & \multicolumn{5}{c}{\textbf{OVERALL}} 
            \\ \cmidrule{2-26}
            & \begin{tabular}{c} Residual \\ Files size\end{tabular} & \begin{tabular}{c} Size of \\ Comm. \end{tabular} & \begin{tabular}{c} Total size \\ of $\mathcal{R}$ \end{tabular} & \begin{tabular}{c} Tot. size \\ Projects \end{tabular} & \% $\mathcal{R}$  
            & \begin{tabular}{c} Residual \\ Files size\end{tabular} & \begin{tabular}{c} Size of \\ Comm. \end{tabular} & \begin{tabular}{c} Total size \\ of $\mathcal{R}$ \end{tabular} & \begin{tabular}{c} Tot. size \\ Projects \end{tabular} & \% $\mathcal{R}$  
            & \begin{tabular}{c} Residual \\ Files size\end{tabular} & \begin{tabular}{c} Size of \\ Comm. \end{tabular} & \begin{tabular}{c} Total size \\ of $\mathcal{R}$ \end{tabular} & \begin{tabular}{c} Tot. size \\ Projects \end{tabular} & \% $\mathcal{R}$  
            & \begin{tabular}{c} Residual \\ Files size\end{tabular} & \begin{tabular}{c} Size of \\ Comm. \end{tabular} & \begin{tabular}{c} Total size \\ of $\mathcal{R}$ \end{tabular} & \begin{tabular}{c} Tot. size \\ Projects \end{tabular} & \% $\mathcal{R}$ 
            & \begin{tabular}{c} Residual \\ Files size\end{tabular} & \begin{tabular}{c} Size of \\ Comm. \end{tabular} & \begin{tabular}{c} Total size \\ of $\mathcal{R}$ \end{tabular} & \begin{tabular}{c} Tot. size \\ Projects \end{tabular} & \% $\mathcal{R}$  \\
            \midrule

            2025 & 24,736 & 162 & 24,898 & 94,718 & 26.29\% & 31,205 & 191 & 31,396 & 112,924 & 27.80\% & 37,484 & 210 & 37,694 & 142,312 & 26.49\% & 32,389 & 183 & 32,572 & 115,694 & 28.15\% & 125,815 & 746 & 126,561 & 465,648 & 27.18\% \\
            2024 & 23,883 & 150 & 24,033 & 88,153 & 27.26\% & 29,472 & 174 & 29,646 & 103,752 & 28.57\% & 31,761 & 175 & 31,936 & 114,050 & 28.00\% & 29,525 & 168 & 29,693 & 105,045 & 28.27\% & 114,641 & 667 & 115,308 &  411,001 & 28.06\% \\
            2023 & 20,542 & 112 & 20,654 & 63,353 & 32.60\% & 21,689 & 126 & 21,815 & 72,129 & 30.24\% & 29,565 & 147 & 29,712 & 94,572 & 31.42\% & 23,328 & 119 & 23,447 & 75,713 & 30.97\% & 95,123 & 504 & 95,627 & 305,768 & 31.27\% \\
            2022 & 17,662 & 101 & 17,763 & 55,346 & 32.09\% & 17,700 & 116 & 17,816 & 57,681 & 30.89\% & 26,108 & 139 & 26,247 & 80,787 & 32.49\% & 20,858 & 113 & 20,971 & 64,034 & 32.75\% & 82,328 & 469 & 82,797 & 257,848 & 32.11\% \\
            2021 & 15,544 & 92 & 15,636 & 48,512 & 32.23\% & 16,329 & 107 & 16,436 & 51,692 & 31.80\% & 22,692 & 132 & 22,824 & 75,185 & 30.36\% & 20,466 & 113 & 20,579 & 65,721 & 31.31\% & 75,031 & 444 & 75,475 & 241,110 & 31.30\% \\
            2020 & 6,690 & 81 & 6,771 & 27,412 & 24.70\% & 7,665 & 94 & 7,759 & 30,378 & 25.54\% & 8,504 & 102 & 8,606 & 35,141 & 24.49\% & 8,933 & 108 & 9,041 & 36,430 & 24.82\% & 31,793 & 385 & 32,178 & 129,362 & 24.87\% \\
            2019 & 4,062 & 74 & 4,136 & 22,562 & 18.33\% & 4,108 & 76 & 4,184 & 21,759 & 19.23\% & 4,383 & 82 & 4,465 & 23,983 & 18.62\% & 5,103 & 88 & 5,191 & 27,347 & 18.98\% & 17,656 & 320 & 17,976 & 95,651 & 18.79\% \\
            2018 & 2,727 & 63 & 2,790 & 18,468 & 15.11\% & 3,165 & 66 & 3,231 & 19,335 & 16.71\% & 3,426 & 68 & 3,494 & 21,597 & 16.18\% & 3,511 & 70 & 3,581 & 21,507 & 16.65\% & 12,828 & 267 & 13,095 & 80,907 & 16.19\% \\
            2017 & 1,855 & 52 & 1,907 & 15,001 & 12.71\% & 1,997 & 53 & 2,050 & 14,466 & 14.17\% & 2,652 & 63 & 2,715 & 18,735 & 14.49\% & 2,446 & 53 & 2,499 & 15,535 & 16.09\% & 8,950 & 221 & 9,171 & 63,737 & 14.39\% \\
            2016 & 2,059 & 42 & 2,101 & 13,901 & 15.11\% & 2,001 & 53 & 2,054 & 14,306 & 14.36\% & 2,373 & 55 & 2,428 & 17,184 & 14.13\% & 2,400 & 50 & 2,450 & 16,139 & 15.18\% & 8,832 & 200 & 9,032 & 61,531 & 14.68\% \\
            2015 & 1,612 & 37 & 1,649 & 12,439 & 13.26\% & 1,743 & 42 & 1,785 & 13,189 & 13.53\% & 2,169 & 45 & 2,214 & 14,615 & 15.15\% & 1,844 & 44 & 1,888 & 14,346 & 13.16\% & 7,367 & 168 & 7,535 & 54,590 & 13.80\% \\ \midrule
            
            (agg) & 121,371 & 966 & 122,337 & 459,866 & 26.60\% & 137,074 & 1,098 & 138,172 & 511,614 & 27.01\% & 171,118 & 1,218 & 172,336 & 638,160 & 27.01\% & 150,802 & 1,109 & 151,911 & 557,512 & 27.25\% & 580,365 & 4,391 & 584,756 & 2,167,153 & 26.98\% \\

            \bottomrule
        \end{tabular}
    }
\end{table*}

\begin{table*}[!htbp]
    \centering
    \caption{\textbf{Distribution of projects according to the total size of their residual files ($\mathcal{F}$).} Note that $\mathcal{R}$=$\mathcal{F}$+Comments } 
    \label{tab:residual_distribution_bytes-months}
    \vspace{-3mm}
    \resizebox{2.1\columnwidth}{!}{
        \begin{tabular}{c||c|c|c?c|c|c?c|c|c?c|c|c?c|c|c}
            \toprule
            \multirow{2}{*}{\textbf{Year}}
            
            & \multicolumn{3}{c?}{\textbf{January}} 
            & \multicolumn{3}{c?}{\textbf{February}} 
            & \multicolumn{3}{c?}{\textbf{March}} 
            & \multicolumn{3}{c?}{\textbf{April}} 
            & \multicolumn{3}{c}{\textbf{OVERALL}} 
            \\ \cmidrule{2-16}
            & $\mathcal{F}$<1KB & 1KB<$\mathcal{F}$<1MB & $\mathcal{F}$>1MB  
            & $\mathcal{F}$<1KB & 1KB<$\mathcal{F}$<1MB & $\mathcal{F}$>1MB  
            & $\mathcal{F}$<1KB & 1KB<$\mathcal{F}$<1MB & $\mathcal{F}$>1MB  
            & $\mathcal{F}$<1KB & 1KB<$\mathcal{F}$<1MB & $\mathcal{F}$>1MB 
            & $\mathcal{F}$<1KB & 1KB<$\mathcal{F}$<1MB & $\mathcal{F}$>1MB  \\
            \midrule
            2025 & 4822 (28.07\%) & 8747 (50.93\%) & 3607 (21.00\%) & 4766 (24.83\%) & 10222 (53.25\%) & 4208 (21.92\%) & 5405 (24.64\%) & 11463 (52.25\%) & 5069 (23.11\%) & 5280 (26.75\%) & 10132 (51.34\%) & 4324 (21.91\%) & 20273 (25.98\%) & 40564 (51.98\%) & 17208 (22.05\%) \\
            2024 & 5005 (31.61\%) & 7670 (48.45\%) & 3157 (19.94\%) & 4901 (28.28\%) & 8771 (50.60\%) & 3661 (21.12\%) & 4999 (27.49\%) & 9069 (49.87\%) & 4116 (22.64\%) & 5074 (28.64\%) & 8900 (50.24\%) & 3741 (21.12\%) & 19979 (28.93\%) & 34410 (49.82\%) & 14675 (21.25\%) \\
            2023 & 4295 (35.45\%) & 5016 (41.40\%) & 2804 (23.14\%) & 4227 (31.91\%) & 5722 (43.19\%) & 3299 (24.90\%) & 5171 (31.79\%) & 7081 (43.54\%) & 4013 (24.67\%) & 4257 (32.11\%) & 5866 (44.25\%) & 3133 (23.63\%) & 17950 (32.71\%) & 23685 (43.15\%) & 13249 (24.14\%) \\
            2022 & 4522 (38.75\%) & 4506 (38.62\%) & 2641 (22.63\%) & 4562 (36.93\%) & 4996 (40.44\%) & 2796 (22.63\%) & 5286 (34.53\%) & 6267 (40.94\%) & 3755 (24.53\%) & 4530 (35.92\%) & 5102 (40.46\%) & 2979 (23.62\%) & 18900 (36.39\%) & 20871 (40.18\%) & 12171 (23.43\%) \\
            2021 & 4638 (42.01\%) & 4115 (37.28\%) & 2286 (20.71\%) & 4805 (40.16\%) & 4640 (38.78\%) & 2520 (21.06\%) & 5892 (38.99\%) & 5804 (38.41\%) & 3415 (22.60\%) & 5227 (39.29\%) & 5206 (39.13\%) & 2870 (21.57\%) & 20562 (39.99\%) & 19765 (38.44\%) & 11091 (21.57\%) \\
            2020 & 5196 (49.69\%) & 3723 (35.61\%) & 1537 (14.70\%) & 5412 (47.30\%) & 4220 (36.88\%) & 1809 (15.81\%) & 5940 (46.93\%) & 4707 (37.19\%) & 2011 (15.89\%) & 6144 (46.61\%) & 4964 (37.66\%) & 2073 (15.73\%) & 22692 (47.54\%) & 17614 (36.90\%) & 7430 (15.56\%) \\
            2019 & 6671 (65.56\%) & 2404 (23.62\%) & 1101 (10.82\%) & 6401 (64.23\%) & 2499 (25.08\%) & 1065 (10.69\%) & 7160 (64.46\%) & 2799 (25.20\%) & 1149 (10.34\%) & 7461 (63.11\%) & 3064 (25.92\%) & 1298 (10.98\%) & 27693 (64.29\%) & 10766 (25.00\%) & 4613 (10.71\%) \\
            2018 & 6730 (71.73\%) & 1944 (20.72\%) & 708 (7.55\%) & 6505 (69.22\%) & 2041 (21.72\%) & 851 (9.06\%) & 7039 (69.47\%) & 2179 (21.51\%) & 914 (9.02\%) & 6988 (69.15\%) & 2182 (21.59\%) & 936 (9.26\%) & 27262 (69.87\%) & 8346 (21.39\%) & 3409 (8.74\%) \\
            2017 & 6180 (76.14\%) & 1416 (17.44\%) & 521 (6.42\%) & 5912 (74.18\%) & 1510 (18.95\%) & 548 (6.88\%) & 7275 (73.63\%) & 1920 (19.43\%) & 685 (6.93\%) & 5819 (72.52\%) & 1511 (18.83\%) & 694 (8.65\%) & 25186 (74.10\%) & 6357 (18.70\%) & 2448 (7.20\%) \\
            2016 & 5727 (77.73\%) & 1221 (16.57\%) & 420 (5.70\%) & 6215 (76.07\%) & 1471 (18.00\%) & 484 (5.92\%) & 6661 (76.06\%) & 1569 (17.92\%) & 527 (6.02\%) & 6059 (75.78\%) & 1443 (18.05\%) & 494 (6.18\%) & 24662 (76.37\%) & 5704 (17.66\%) & 1925 (5.96\%) \\
            2015 & 5542 (79.02\%) & 1144 (16.31\%) & 327 (4.66\%) & 5613 (78.73\%) & 1126 (15.79\%) & 390 (5.47\%) & 6401 (78.85\%) & 1278 (15.74\%) & 439 (5.41\%) & 5914 (78.95\%) & 1219 (16.27\%) & 358 (4.78\%) & 23470 (78.89\%) & 4767 (16.02\%) & 1514 (5.09\%) \\ \midrule
            
            (agg) & 59328 (49.30\%) & 41906 (34.82\%) & 19109 (15.88\%) & 59319 (46.28\%) & 47218 (36.84\%) & 21631 (16.88\%) & 67229 (45.59\%) & 54136 (36.71\%) & 26093 (17.70\%) & 62753 (46.40\%) & 49589 (36.67\%) & 22900 (16.93\%) & 248629 (46.80\%) & 192849 (36.30\%) & 89733 (16.89\%) \\

            \bottomrule
        \end{tabular}
    }
\end{table*}

\begin{table*}[!htbp]
    \centering
    \caption{\textbf{Distribution of projects according to the ratio: (total size of residual files)/(total project size).} $\mathcal{F}$=``size of residual files''.} 
    \label{tab:residual_distribution_relative-months}
    \vspace{-3mm}
    \resizebox{2.1\columnwidth}{!}{
        \begin{tabular}{c||c|c|c|c?c|c|c|c?c|c|c|c?c|c|c|c?c|c|c|c}
            \toprule

            \multirow{2}{*}{\textbf{Year}}
            
            & \multicolumn{4}{c?}{\textbf{January}} 
            & \multicolumn{4}{c?}{\textbf{February}} 
            & \multicolumn{4}{c?}{\textbf{March}} 
            & \multicolumn{4}{c?}{\textbf{April}} 
            & \multicolumn{4}{c}{\textbf{OVERALL}} 
            \\ \cmidrule{2-21}
            & $\mathcal{F} <$ 5\% & 5\%$\leq \mathcal{F} <$50\% & 50\%$\leq \mathcal{F} <$95\% & $\mathcal{F}\geq$95\%   
            & $\mathcal{F} <$ 5\% & 5\%$\leq \mathcal{F} <$50\% & 50\%$\leq \mathcal{F} <$95\% & $\mathcal{F}\geq$95\% 
            & $\mathcal{F} <$ 5\% & 5\%$\leq \mathcal{F} <$50\% & 50\%$\leq \mathcal{F} <$95\% & $\mathcal{F}\geq$95\% 
            & $\mathcal{F} <$ 5\% & 5\%$\leq \mathcal{F} <$50\% & 50\%$\leq \mathcal{F} <$95\% & $\mathcal{F}\geq$95\% 
            & $\mathcal{F} <$ 5\% & 5\%$\leq \mathcal{F} <$50\% & 50\%$\leq \mathcal{F} <$95\% & $\mathcal{F}\geq$95\% \\
            \midrule
            2025 & 9275 (54.00\%) & 5421 (31.56\%) & 2375 (13.83\%) & 105 (0.61\%) & 9953 (51.85\%) & 6369 (33.18\%) & 2713 (14.13\%) & 161 (0.84\%) & 11616 (52.95\%) & 7100 (32.37\%) & 3080 (14.04\%) & 141 (0.64\%) & 10493 (53.17\%) & 6177 (31.30\%) & 2920 (14.80\%) & 146 (0.74\%) & 41337 (52.97\%) & 25067 (32.12\%) & 11088 (14.21\%) & 553 (0.71\%) \\
            2024 & 8849 (55.89\%) & 4733 (29.90\%) & 2098 (13.25\%) & 152 (0.96\%) & 9314 (53.74\%) & 5443 (31.40\%) & 2363 (13.63\%) & 213 (1.23\%) & 9677 (53.22\%) & 5607 (30.83\%) & 2736 (15.05\%) & 164 (0.90\%) & 9589 (54.13\%) & 5517 (31.14\%) & 2441 (13.78\%) & 168 (0.95\%) & 37429 (54.19\%) & 21300 (30.84\%) & 9638 (13.96\%) & 697 (1.01\%) \\
            2023 & 6706 (55.35\%) & 3205 (26.45\%) & 2088 (17.23\%) & 116 (0.96\%) & 6920 (52.23\%) & 3750 (28.31\%) & 2464 (18.60\%) & 114 (0.86\%) & 8628 (53.05\%) & 4549 (27.97\%) & 2955 (18.17\%) & 133 (0.82\%) & 7222 (54.48\%) & 3594 (27.11\%) & 2318 (17.49\%) & 122 (0.92\%) & 29476 (53.71\%) & 15098 (27.51\%) & 9825 (17.90\%) & 485 (0.88\%) \\
            2022 & 6636 (56.87\%) & 2887 (24.74\%) & 2047 (17.54\%) & 99 (0.85\%) & 6863 (55.55\%) & 3283 (26.57\%) & 2110 (17.08\%) & 98 (0.79\%) & 8267 (54.00\%) & 4096 (26.76\%) & 2820 (18.42\%) & 125 (0.82\%) & 6907 (54.77\%) & 3231 (25.62\%) & 2357 (18.69\%) & 116 (0.92\%) & 28673 (55.20\%) & 13497 (25.98\%) & 9334 (17.97\%) & 438 (0.84\%) \\
            2021 & 6566 (59.48\%) & 2656 (24.06\%) & 1705 (15.45\%) & 112 (1.01\%) & 6884 (57.53\%) & 2966 (24.79\%) & 2004 (16.75\%) & 111 (0.93\%) & 8562 (56.66\%) & 3813 (25.23\%) & 2634 (17.43\%) & 102 (0.68\%) & 7694 (57.84\%) & 3221 (24.21\%) & 2268 (17.05\%) & 120 (0.90\%) & 29706 (57.77\%) & 12656 (24.61\%) & 8611 (16.75\%) & 445 (0.87\%) \\
            2020 & 6926 (66.24\%) & 2151 (20.57\%) & 1316 (12.59\%) & 63 (0.60\%) & 7271 (63.55\%) & 2631 (23.00\%) & 1483 (12.96\%) & 56 (0.49\%) & 8084 (63.86\%) & 2907 (22.97\%) & 1593 (12.58\%) & 74 (0.58\%) & 8417 (63.86\%) & 3000 (22.76\%) & 1693 (12.84\%) & 71 (0.54\%) & 30698 (64.31\%) & 10689 (22.39\%) & 6085 (12.75\%) & 264 (0.55\%) \\
            2019 & 7514 (73.84\%) & 1677 (16.48\%) & 935 (9.19\%) & 50 (0.49\%) & 7354 (73.80\%) & 1638 (16.44\%) & 906 (9.09\%) & 67 (0.67\%) & 8211 (73.92\%) & 1859 (16.74\%) & 991 (8.92\%) & 47 (0.42\%) & 8608 (72.81\%) & 2129 (18.01\%) & 1032 (8.73\%) & 54 (0.46\%) & 31687 (73.57\%) & 7303 (16.96\%) & 3864 (8.97\%) & 218 (0.51\%) \\
            2018 & 7425 (79.14\%) & 1256 (13.39\%) & 650 (6.93\%) & 51 (0.54\%) & 7256 (77.22\%) & 1296 (13.79\%) & 778 (8.28\%) & 67 (0.71\%) & 7823 (77.21\%) & 1455 (14.36\%) & 802 (7.92\%) & 52 (0.51\%) & 7774 (76.92\%) & 1468 (14.53\%) & 799 (7.91\%) & 65 (0.64\%) & 30278 (77.60\%) & 5475 (14.03\%) & 3029 (7.76\%) & 235 (0.60\%) \\
            2017 & 6722 (82.81\%) & 871 (10.73\%) & 470 (5.79\%) & 54 (0.67\%) & 6482 (81.33\%) & 934 (11.72\%) & 501 (6.29\%) & 53 (0.66\%) & 7976 (80.73\%) & 1186 (12.00\%) & 654 (6.62\%) & 64 (0.65\%) & 6398 (79.74\%) & 966 (12.04\%) & 602 (7.50\%) & 58 (0.72\%) & 27578 (81.13\%) & 3957 (11.64\%) & 2227 (6.55\%) & 229 (0.67\%) \\
            2016 & 6197 (84.11\%) & 711 (9.65\%) & 405 (5.50\%) & 55 (0.75\%) & 6746 (82.57\%) & 876 (10.72\%) & 488 (5.97\%) & 60 (0.73\%) & 7297 (83.33\%) & 910 (10.39\%) & 500 (5.71\%) & 50 (0.57\%) & 6635 (82.98\%) & 856 (10.71\%) & 448 (5.60\%) & 57 (0.71\%) & 26875 (83.23\%) & 3353 (10.38\%) & 1841 (5.70\%) & 222 (0.69\%) \\
            2015 & 5993 (85.46\%) & 621 (8.85\%) & 347 (4.95\%) & 52 (0.74\%) & 6048 (84.84\%) & 655 (9.19\%) & 354 (4.97\%) & 72 (1.01\%) & 6944 (85.54\%) & 679 (8.36\%) & 428 (5.27\%) & 67 (0.83\%) & 6381 (85.18\%) & 682 (9.10\%) & 374 (4.99\%) & 54 (0.72\%) & 25366 (85.26\%) & 2637 (8.86\%) & 1503 (5.05\%) & 245 (0.82\%) \\ \midrule
            
            (agg) & 78809 (65.49\%) & 26189 (21.76\%) & 14436 (12.00\%) & 909 (0.76\%) & 81091 (63.27\%) & 29841 (23.28\%) & 16164 (12.61\%) & 1072 (0.84\%) & 93085 (63.13\%) & 34161 (23.17\%) & 19193 (13.02\%) & 1019 (0.69\%) & 86118 (63.68\%) & 30841 (22.80\%) & 17252 (12.76\%) & 1031 (0.76\%) & 339103 (63.84\%) & 121032 (22.78\%) & 67045 (12.62\%) & 4031 (0.76\%) \\

            \bottomrule
        \end{tabular}
    }
\end{table*}

\begin{figure*}
    \centering
    \begin{subfigure}[t]{0.49\textwidth}
        \centering
        \includegraphics[width=\columnwidth]{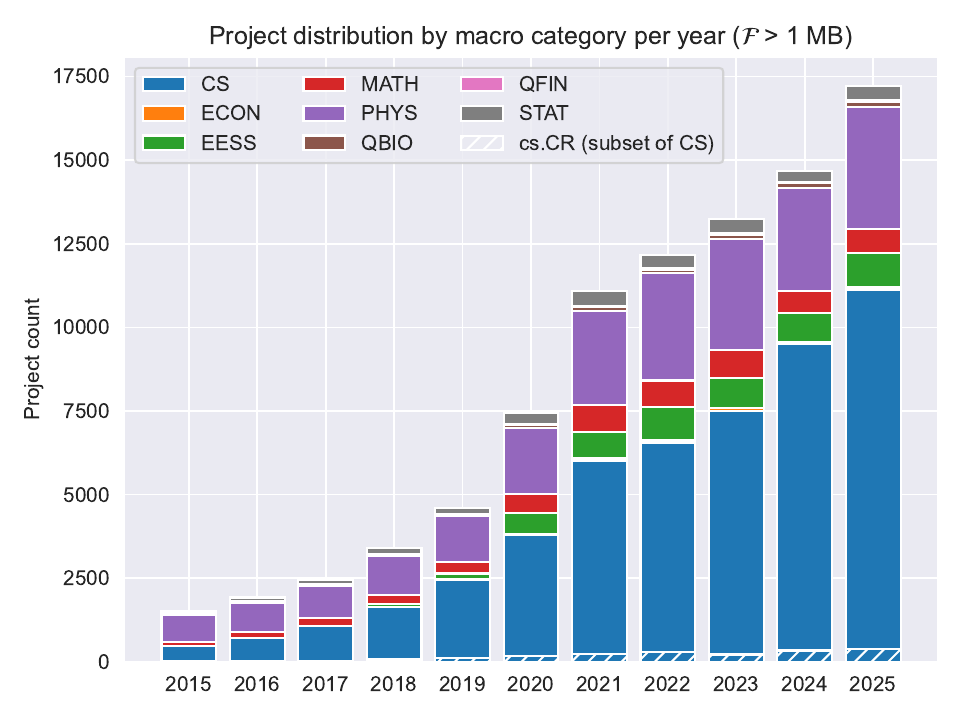}
    \end{subfigure}%
    ~ 
    \begin{subfigure}[t]{0.49\textwidth}
        \centering
        \includegraphics[width=\columnwidth]{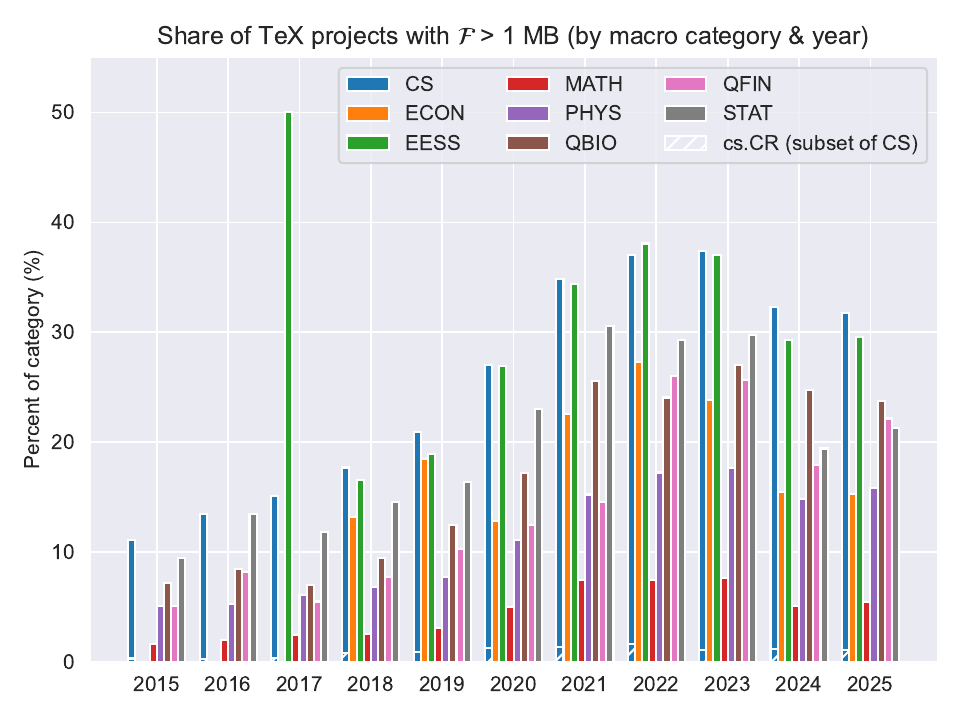}
    \end{subfigure}
    \vspace{-3mm}
    \caption{Distribution (absolute on the left, relative on the right) of projects with >1MB of residual files across scientific categories.}
    \label{fig:onemb_categories}
\end{figure*}

\begin{figure}[!t]
    \centering
    \includegraphics[width=\columnwidth]{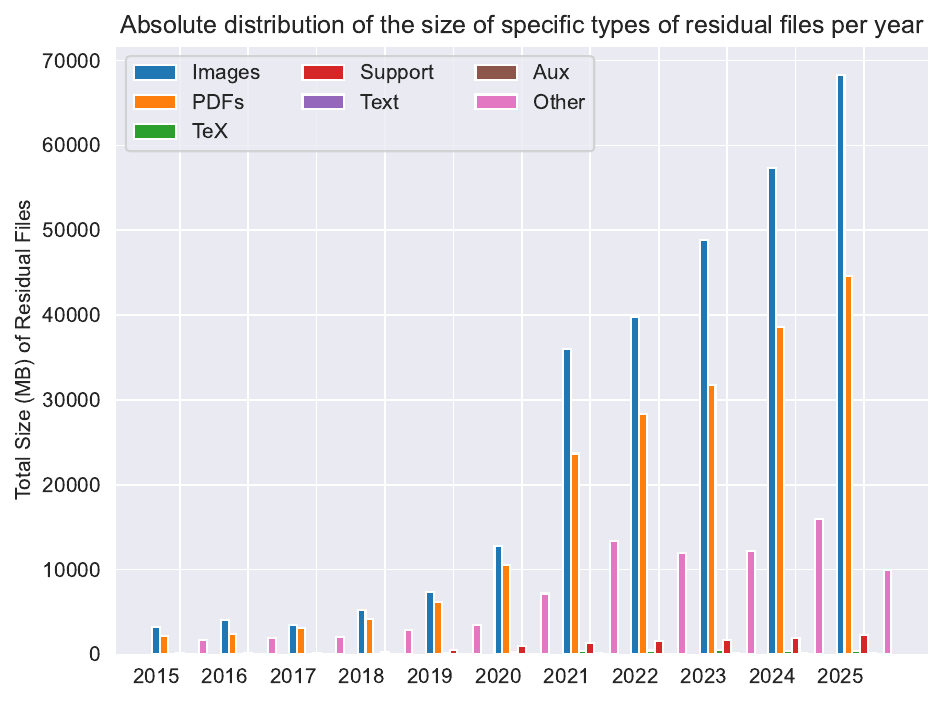}
    \vspace{-5mm}
    \caption{Total size of file types across the residual files.}
    \label{fig:types_size}
    \vspace{-3mm}
\end{figure}
\section{Additional details on \texttt{BaRDE}}
\label{app:barde}
\noindent
We report additional low-level information on \texttt{BaRDE}.

\subsection{Functions}
\label{sapp:functions}

\noindent
We report in Alg.~\ref{alg:ext} some extended function os \texttt{BaRDE}. Note: the functions extractComments(\smamath{file}), applyHeuristics(\smamath{content}), and findResidual(\smamath{root,unpacked}) are discussed in §\ref{sssec:heuristics}. 

\begin{algorithm2e}
    \setcounter{AlgoLine}{23}
    \caption{Functions of \texttt{BaRDE} (extending Algorithm~\ref{alg:barde})}
    \footnotesize
    \label{alg:ext}
    \DontPrintSemicolon
    \SetAlgoNoEnd

    {\tt \scriptsize // Procedure for submissions consisting in a single TeX file}\\

    \SetKwFunction{proc}{proc}
    \SetKwProg{myproc}{Function}{}{}
    \myproc{singleTeX($file, submission.ID$)}{ 
    \label{alg:singletex}
        $used, residual \gets$ emptyList();\\
        Insert $file.name$ in $used$;\\
        $statistics \gets$ computeStats($used, residual$);\\
        $comments \gets$ extractComments($file$);\\
        \Return ($submission.ID, used, residual, statistics, comments$);\\
    }

    {\tt \scriptsize // Procedure that determines the size of used and unused files of a TeX project}\\

    \SetKwFunction{proc}{proc}
    \SetKwProg{myproc}{Function}{}{}
    \myproc{computeStats($used, residual$)}{ 
    \label{alg:computestats}
        $used\_sizes, residual\_sizes$ = 0;\\
        \For{file \KwOf used}{
            Insert $file.size$ in $used\_sizes$
        }
        \For{file \KwOf residual}{
            Insert $file.size$ in $residual\_sizes$
        }
        \Return $used\_sizes, residual\_sizes$
    }

    {\tt \scriptsize // Procedure that tries to find the root TeX file in a blob, returning $\emptyset$ if not successful}\\

    \SetKwFunction{proc}{proc}
    \SetKwProg{myproc}{Function}{}{}
    \myproc{inferRoot($blob$)}{ 
    \label{alg:inferRoot}
        $content \gets$ unpack($blob$);\\
        $root \gets$ applyHeuristics($content$);\\
        \Return $root, content$
    }
    
    {\tt \scriptsize // Procedure for full-fledged TeX projects}\\

    \SetKwFunction{proc}{proc}
    \SetKwProg{myproc}{Function}{}{}
    \myproc{multiTeX($root, unpacked, submission.ID$)}{ 
    \label{alg:multitex}
        $used, residual \gets$ findResidual($root, unpacked$);\\
        $statistics \gets$ computeStats($used, residual$);\\
        $comments \gets \emptyset$ ;\\
        \For{file $\KwOf$ used $\KwIf$ file.type="TeX"}{
            $comments$ += extractComments($file$);\\
        }
        \Return ($submission.ID, used, residual, statistics, comments$);\\
    }
\end{algorithm2e}

\subsection{Patterns and Special cases}
\label{sapp:patterns}

\noindent
We report in Listing~\ref{lst:patterns} the list of commands that \texttt{BaRDE} considers when searching for ``used'' files when scanning a {\footnotesize \TeX{}} project. These commands are either well-known, or we encountered them while troubleshooting some projects while developing \texttt{BaRDE}. Note that, in our implementation, each command is provided with a specific regex capture pattern, which enables to also cover cases in which there are additional options specified before the filename is provided. For instance, the capture pattern for {\small \texttt{\textbackslash includegraphics}} is: 

{\footnotesize
\begin{verbatim}
("includegraphics", re.compile(r'\\includegraphics\*?
\s*(?:\[[^\]]*\])?\s*\{+\s*([^{}]+?)\s*\}+', re.DOTALL))
\end{verbatim}
}

We also implemented handling of potential exceptions. 
When looking if a file ``caught'' in any pattern is present in the project, we first consider looking for it in the directory of the file being analyzed; if no match is found, we look in the topmost folder of the project; and we also consider looking in a potential {\ttfamily\small graphicspath} (if there is a mention). We also handled cases mentioning {\ttfamily\small pdfmapfile}. Moreover, {\footnotesize \TeX{}} may automatically append the extension of a file when certain methods are invoked (e.g., using {\ttfamily\small \textbackslash input\{main\}} or {\ttfamily\small \textbackslash input\{main.tex\}} is equivalent), so when doing the matching between the files `seen' and those included in the project, we also try extensions that typically align with that of the command being invoked (e.g., for {\small \texttt{\textbackslash includegraphics}}, we consider: .png, .svg, .jpg, .pdf, .jpeg, .eps, .svg) and, in other cases, we try a list of well-known extensions (see Listing~\ref{lst:extensions}). We also accounted for case-sensitivity, folder separators, whitespaces, preliminary options, quotation marks.

\begin{lstlisting}[
frame=single,
breaklines=true, 
mathescape=false,
basicstyle=\ttfamily\footnotesize,
caption=Commands to consider when finding ``used'' files.,
label={lst:patterns},
basicstyle=\scriptsize,
belowskip=-5mm,
float=t
]

input, include, includegraphics, includesvg, overpic, pgfimage, epsfig, pdfximage, bibliography, usepackage, documentclass, includepdf, pdfmapfile, pdfmapline, requirepackage, LoadClass, addbibresource, lstinputlisting, inputminted, verbatiminput, includeonly, subimport, includefrom, subincludefrom, subfile, includestandalone, externaldocument, usetikzlibrary, usepgfplotslibrary, pgfdeclareimage, pgfplotstableread, addplot_table, addplot_graphics, csvautotabular, csvreader, DTLloaddb, SweaveInput, bibliographystyle, movie, readdef, loadglsentries, InputIfFileExists, DeclareFontShape, plotone, plottwo, plotfiddle, tikzfig, trimfig, biographywithpic
  
\end{lstlisting}

\begin{lstlisting}[
frame=single,
breaklines=true, 
mathescape=false,
basicstyle=\ttfamily\footnotesize,
caption=Extensions added when looking for used files and extraction of font files.,
label={lst:extensions},
basicstyle=\scriptsize,
belowskip=-5mm,
float=t
]

exts_to_try = ['', '.tex', '.pdf', '.png', '.jpg', '.jpeg', '.eps', '.svg', '.bmp', '.sty', '.cls', '.bib']

font_file_pattern = re.compile(r'<(?:\[)?([^<>\[\]\s]+?\.(?:tfm|vf|pfb|enc|fd))')
  
\end{lstlisting}

\subsection{Type of files}
\label{sapp:type}
\noindent
When determining the specific type of an (unused) file, we inspected its extension and assigned it to a specific group. We report in Listing~\ref{lst:types} the extensions associated to each group. Any file with an extension not mentioned in Listing~\ref{lst:types} is considered as ``other''.

\begin{lstlisting}[
frame=single,
breaklines=true, 
mathescape=false,
basicstyle=\ttfamily\footnotesize,
caption=Type of files according to their extension.,
label={lst:types},
basicstyle=\scriptsize,
belowskip=-5mm,
float=t
]

image_exts = {'.png', '.jpg', '.jpeg', '.eps', '.svg', '.bmp', '.tiff', '.tif', '.gif'}
pdf_exts = {'.pdf'}
tex_exts = {'.tex'}
support_exts = {'.sty', '.cls', '.bst', '.bbl', '.aux', '.lof', '.lot', '.out', '.toc', '.synctex.gz', '.fls', '.fdb_latexmk'}
aux_exts = {'.tfm', '.vf', '.fd', '.pfb', '.map', '.enc'}
text_exts = {'.txt', '.md'}
  
\end{lstlisting}

\section{Considerations on our Results}
\label{sec:considerations}
\noindent
We extend our findings (in both §\ref{sec:residual} and §\ref{sec:problematic}) with additional considerations. First, we discuss some outliers~(Appendix~\ref{ssec:outlier}). Then, we discuss some lessons learned~(Appendix~\ref{ssec:ancillary}). Finally, we compare \texttt{BaRDE} with ALC~(Appendix~\ref{ssec:comparison}).

\subsection{Outlier analysis}
\label{ssec:outlier}
\noindent
The quantitative results we presented (in §\ref{sec:residual}) have been derived after running \texttt{BaRDE} on our dataset multiple times---each time leading to an improvement of \texttt{BaRDE}.

Indeed, whenever we finished a single run of \texttt{BaRDE}, we (manually) inspected some ``outliers'' to determine if the residual files found by by \texttt{BaRDE} were truly not necessary to make the PDF: whenever we found an error, typically due to pathing\,/\,naming issues, or due to not including some methods used by {\footnotesize \TeX{}} to reference other files in a project (e.g., we were initially not aware of the {\small \texttt{\textbackslash overpic}} method), we revised \texttt{BaRDE}'s source code, and ran it again on our dataset. 

Let us provide some examples of ``noteworthy outliers'' (which we manually checked):
\begin{itemize}[leftmargin=*]
    \item We looked at the outliers in the {\footnotesize \TeX{}} file group. Some projects had residual {\footnotesize \TeX{}} files of >\,20MB in size, representing code for Ti\textit{k}z~\cite{tikz} pictures that were not included in the paper;
    \item We looked at the outliers in the ``support'' category: one submission had over 20\,MBs of .out files containing experimental results (and also revealing other sensitive details on the user\,/\,platform used for the experiments);
    \item We looked at the outliers in the ``other'' category, and we found projects with 100s-of-MBs worth of video files;
    \item One submission had over 1,700 unused PDF images.
\end{itemize}
The submission with the highest number of residual files had 7,589 residual files (out of 7,744).

\subsection{Ancillary Findings}
\label{ssec:ancillary}
\noindent
In the course of our research, we encountered a variety of intriguing special cases that we believe should be reported.
\begin{itemize}[leftmargin=*]
    \item For some venues (e.g., AAAI'25~\cite{aaai25_template}), the {\footnotesize \TeX{}} template for the camera-ready of accepted papers explicitly states ``{\small \textsf{Do not send files that are not actually used in the paper}}''. We found interesting that such lines were included in the comments of submissions for which we found residual files;
    \item We found many papers using the ``math commands'' template files created by Ian Goodfellow~\cite{mathcommands};
    \item We excluded one project because it kept \textit{triggering our antivirus} (Microsoft Defender, updated in July 2025).
\end{itemize}
We also make two noteworthy considerations.

First, \textbf{smaller is not safer}. We found potentially-sensitive data even in projects with just a few residual files (e.g., 2 or 3) of relatively modest size (<5\% of the project size). Hence, even though submissions with a low ratio, or low overall size, of residual data may be less likely to contain sensitive data, it would be misleading (and dangerous) to focus only on submissions with considerable amounts of residual data.

Second, \textbf{non-residual files can be problematic, too}. A submission in its \textit{used files} had offensive names (in this case, a used image named ``{\small \textsf{diagram\_fuck}}'' (and, after manually checking the paper, we convened it had nothing to do with ``fuck''). Hence, even by removing all residual data, source files can still include potentially-sensitive elements that the authors should be aware of before making the submission.

\subsection{Comparison with arxiv-latex-cleaner}
\label{ssec:comparison}

\noindent
The arxiv-latex-cleaner, or ALC for short, is a tool designed to remove unnecessary data from a given {\footnotesize \TeX{}} project meant to be submitted to arXiv~\cite{arxiv_cleaner}. Despite its relatively widespread usage (the ALC repository has 6.2k stars on GitHub), ALC is not suited for the analysis carried out in this paper. 

This unsuitability is due to ALC having a ``project-specific'' focus. The idea is that authors, who have complete knowledge of their {\footnotesize \TeX{}} projects, use ALC to, e.g., save space, or remove comments. However, for our analysis, we {\small \textit{(i)}}~must inspect thousands of projects and {\small \textit{(ii)}}~we do not know what each project contains. Hence, ALC has three noteworthy limitations that prevent its usage for the sake of our research. 
\begin{itemize}[leftmargin=*]
    \item \textbf{No support for bulk analyses.} ALC is designed to simply ``clean'' a single and user-provided {\footnotesize \TeX{}} project from files superfluous for arXiv. Therefore, ALC does not natively support analyses of multiple projects, does not compute statistics on the size of the cleaned data, and does not extract the textual comments for follow-up analyses. All these features are, in contrast, an integral part of \texttt{BaRDE}. 
    
    \item \textbf{No support for GZ or blob files.} ALC accepts a \smamath{project folder} as input, which ideally contains the content of the entire {\footnotesize \TeX{}} project. However, this is not the case for the arXiv submissions' source files we downloaded from AWS S3 which are provided either as PDF or as compressed GZ files---the latter mostly containing packed blob files (see §\ref{ssec:organization}). To provide an idea, running ALC on the files shown in Figure~\ref{fig:subs} always returns errors. Hence, to use ALC on our dataset, it would have been needed to unzip all GZ files, and unpack all the blob files contained therein. These operations are automatically handled by \texttt{BaRDE} at runtime. 
    
    \item \textbf{Simplistic assumption.} ALC assumes the user-provided \smamath{project folder} to have a single {\footnotesize \TeX{}} file -- which is \textit{assumed to be the {\footnotesize \TeX{}} root} -- in its topmost level and then considers all files called by it to build a tree of ``used files''. However, such a logic does not align with the submissions on arXiv, which can have {\small \textit{(a)}}~no {\footnotesize \TeX{}} files in their root---in which case, no tree would be built, leading to erroneously considering all files of the project as ``unused files''; or {\small \textit{(b)}}~multiple {\footnotesize \TeX{}} files in their root---in which case, all such TeX files would be considered as ``root'', leading to false positives.\footnote{For instance, a project may have two {\scriptsize \TeX{}} files in its main folder (e.g., the actual root {\scriptsize \TeX{}} file, as well as another {\scriptsize \TeX{}} file which could be that of a template, or of supplementary material, not called by the root {\scriptsize \TeX{}}): according to ALC, all files called by such {\scriptsize \TeX{}} files would be considered as part of the project, even though they are not part of the final paper produced as a result of the {\scriptsize \TeX{}} compilation. We empirically verified these claims.} In general, ALC does not have any built-in mechanism to infer the root, because ALC expects the project to be in a fixed format---which is not the case for arXiv's submissions. In contrast, \texttt{BaRDE} is designed to handle these cases.
\end{itemize}
Put simply, it is not possible to use ALC for our analyses, not even by ``extending'' ALC with some custom wrapper function. This is why we developed \texttt{BaRDE} from scratch. 

However, we acknowledge that some low-level details of \texttt{BaRDE} have been inspired by ALC, such as what can be considered as a comment or the file-extensions that are not needed by arXiv, which makes \texttt{BaRDE} more robust.

\section{Outreach}
\label{app:communication}
\noindent
We report the email we sent to arXiv's leadership team~\cite{arxiv_leadership} (Email~\ref{emailbox:arxiv}; note that, after sending it, we re-did our analysis after updating \texttt{BaRDE}, which is why the numbers in this email are slightly different from those in this paper), and the email we sent to the authors of submissions with ``problematic'' residual data (Email~\ref{emailbox:authors}).

\vspace{1mm}

\begin{emailBox}{Concerning findings about arXiv}\label{emailbox:arxiv}
{\scriptsize
Dear members of the arXiv Leadership Team,\\

We are [REDACTED]. We are reaching out to you because we’ve discovered some concerning findings related to arXiv’s platform as a whole—findings which you should be aware of.\\

You probably know that, to make a submission on arXiv, it is necessary to provide the source TeX files. You’re probably also aware that, once a submission has been accepted and becomes available on arXiv, anyone can publicly download its corresponding TeX source files. Finally, you’re probably also aware that submitters may not always “clean” their source TeX files of data that is not necessary to produce the corresponding PDF.\\

We have analysed the source files of 600 thousands arXiv submissions. Specifically, we analysed \textit{all} submissions made within the first four months in the 2015–2025 timespan (11 years). Our goal was twofold: (1) measure how much data included in an arXiv submission is not needed to produce its corresponding PDF; and (2) investigate what can be found in such “superfluous” data—which is publicly available.\\

We believe our findings are concerning. First, we found that, overall, there are over 600GB of “useless” data (and since we only considered the first third of every year between 2015–2025, it is safe to assume that the total amount of “useless data” can amount to over 2TB of space that is burdening your servers/storage). Second, we found that, overall, 30\% of the data included in a submission is not necessary to produce its PDF---worryingly, this is an increasing trend (e.g., in 2015 the percentage was only 15\%, whereas it is above 30\% since 2022). Third, we found that such “useless data” contains a variety of sensitive information that we have reason to believe the authors did not intend to make publicly available: for instance, we found instances of offensive language (“WTF does this mean?”), sometimes against other authors/papers (“this stupid paper”); instances of private comments between authors, occasionally suggesting to conceal limitations; as well as instances of undisclosed research data (including links to spreadsheets “open to anybody with the link”) which can lead to scooping; and also malware payload that triggered our antiviruses.\\

Our stance is that authors that submit papers to arXiv are NOT aware that the source files are publicly available. In the sake of responsible disclosure, we are reaching out to you first. However, we are also reaching out to some of the authors of submissions with clearly sensitive data. But of course, we cannot do so for all cases. Therefore, we intend to make our findings public—but no sooner than 3 months from now.
We are available to provide more details on our findings and analytical process. We believe that making submitters more aware of the “public availability” of a submission’s source files would benefit both arXiv (since less space will be used) as well as its users (since no sensitive data would be included—hopefully!).\\

Best regards,
[REDACTED]}
\end{emailBox}
\vspace{-1mm}
{\noindent\small\textbf{Email~\ref{emailbox:arxiv}.} The email we sent to the arXiv leadership team~\cite{arxiv_leadership}.}
\vspace{5mm}

\begin{emailBox}{About your arXiv submission \$submissionID}\label{emailbox:authors}
{\scriptsize
Hello! 
 
We are [REDACTED].\\
 
You are receiving this email because you have submitted a paper on arXiv. Perhaps you were not aware of this, but the source files uploaded on arXiv to create the paper's PDF are public.\\ 
 
While carrying out our research, we found that your submission https://arxiv.org/abs/\$submissionID contains some data (e.g., textual comments, old versions of certain sections, unused figures, undisclosed/confidential research data, or files unrelated to your paper PDF) which you may not have wanted to be publicly available. \\
 
Given this, we have two questions for you, and one recommendation:
\begin{itemize}
    \item \textbf{Q1}: Did you know that all source files uploaded on arXiv to generate the PDF are indeed publicly available?,
    \item \textbf{Q2}: When submitting your paper, was it your intention to include all data currently contained in your source files (including, e.g., textual comments, or unused files)?,
    \item \textbf{Recommendation}: If you were not aware of the "public availability" of a submission's source files, and/or if it was not your intention to include all the data currently contained in your submission's source files, then we want you to know that you CAN UPDATE your source files. Indeed, arXiv only allows downloading the most recent version of a submission's source files, meaning that if you update the source files now, everything currently stored on arXiv would be overwritten. Note, however, that any third-party user who downloaded your submission's source files would still have access to them.\\
\end{itemize}
 
Upon your explicit request, we can delete the source files of your submission that we downloaded for our research. If you want this, we ask that you update your arXiv submission and let us know once the new version is available.\\ 
 
We will then replace your submission's source files we previously downloaded with the updated version available on arXiv.\\
 
We hope that this email was helpful to you. If you have questions, just ask: we'll gladly reply!\\
 
[REDACTED]}
\end{emailBox}
\vspace{-1mm}
{\noindent\small\textbf{Email~\ref{emailbox:authors}.} The email we sent to the authors of ``problematic'' projects.}
\vspace{2mm}

\section{Problematic Comments}
\label{app:excerpts}
\noindent
We provide more details on our codebook, and then provide some excerpts taken from comments. 

Importantly: to protect authors, we \textit{anonymised} specific parts of comments that could be used to identify the submission. Such parts have been replaced with the string: \anonym{} (such a string can denote either a single word, a symbol, a macro, a URL, or multiple words, each of variable length).

\subsection{Motivation for our codebook's codes}
\label{sapp:codebook}
\noindent
Let us provide some examples of how each code in our codebook can be used to cause harm against the authors of a submission presenting such a code: 
\begin{itemize}[leftmargin=*]
    \item \textit{Author Exchange:} these exchanges can be very long, and typically include some information that allows to identify the author of the comment. For instance, one can profile a certain individual through such information (especially because one may ``stalk'' a given author by looking at all the comments left in the arXiv submissions they co-authored). Moreover, such exchanges may reveal, e.g., how a certain academic interacts with other. Taken out of context, such comments may backfire.
    \item \textit{Direct translation:} these can cause concerns because may denote poor knowledge of the English language. For instance, young scholars may be penalized because a given entity finds out that the text in their arXiv submission appears to be written via automated tools.
    \item \textit{Inappropriate language:} depending on the context, consequences may be severe. For instance, using derogatory language against certain scholars may lead to controversies. Whereas using offensive terms can also backfire in professional contexts.
    \item \textit{Data leak:} this can lead to stealing sensitive data (e.g., undisclosed research results), or to setup attacks (e.g., by inferring the software used by an author).
\end{itemize}
For ``other'', circumstances vary. For instance, the presence of a commented-out acknowledgment can raise question of why it was omitted: do the authors plan to show it at a later stage? was it a genuine mistake? was it from a prior template? or was it a deliberate omission because the authors felt that mentioning the acknowledgment may have questioned some of the paper's contributions?

\subsection{Excerpts from the random search}
\label{sapp:excerpts_random}
\noindent
We report 11 excerpts (Excerpt~\ref{exc:one} to~\ref{exc:eleven}) we deemed as ``problematic'' during our random sampling-based manual check. We also report the specific code we assigned to each excerpt (in the heading).

Note that for Excerpt~\ref{exc:four}, the paper never mentions release of code (either before or ``after publication''); whereas for Excerpt~\ref{exc:eleven}, there is no acknowledgement mentioned in the paper (and the entities being acknowledged are not mentioned in the paper).

\begin{excerpt}[label={exc:one}]{Other Problematic}{}
The following is a general outline of a survey paper.\\
Introduction - with background information on the topic and research questions\\
Literature Overview - including relevant research studies and their analysis\\
Methodologies and Approaches - detailing the methods used to collect and analyze data in the literature overview\\
Findings and Trends - summarizing the key findings and trends from the literature review\\
Challenges and Gaps - highlighting the limitations of studies reviewed\\
Future Research Direction - exploring future research opportunities and recommendations\\
Conclusion - a summary of the research conducted and its significance, along with suggestions for further work in this area.\\
References - a list of all the sources cited in the paper, including academic articles and reports.\\
You can always customize this outline to fit your paper's specific requirements, but none of the components can be eliminated. Our custom essay writer\\
Source: https://essaypro.com/blog/how-to-write-a-survey-paper-brief-overview
\end{excerpt}

\begin{excerpt}[label={exc:two}]{Other Problematic}{}
Add 22 more references from European and American authors.
\end{excerpt}

\begin{excerpt}[label={exc:three}]{Author Exchange}{}
\texttt{\textbackslash hl\{}@\anonym{}, what you described here is WHAT tool you used in the implementation but not HOW you implement the system.\texttt{\}}\\
\texttt{\textbackslash hl\{}2. We design a latency reduction strategy that can accelerate the execution of on-device VLMs by XXXXX\texttt{\}}\\
\texttt{\textbackslash hl\{}Recap the observations here: When try to limit the number of gpu in \anonym, \anonym may use the stratgey to allocat part memory in gpu and cpu. but it may be little used with the gpu, even slower than without gpu.\texttt{\}}
\end{excerpt}

\begin{excerpt}[label={exc:four}]{Other Problematic}{}
Upon publication, we will open-source our implementation and make it available for all \anonym{} sensors.
We open-source our code and include it as a standard option for all \anonym{} sensors at \anonym{}.
\end{excerpt}

\begin{excerpt}[label={exc:five}]{Author Exchanges}{}
\texttt{\textbackslash \anonym{}\{}motivate the problem. Explain the problem using the sentence: the dog broke the vase. It was clumsy.\texttt{\}}

\texttt{\textbackslash \anonym{}\{}the diagram doesn't make sense\texttt{\}}

\texttt{\textbackslash \anonym{}\{}diagram of prob dist\texttt{\}}

\texttt{\textbackslash \anonym{}\{}look at the explicit way of mathematically representing this morphism\texttt{\}}

\texttt{\textbackslash \anonym{}\{}this has to be redone completely\texttt{\}}

\%\%\%\%\%\%\%\%\%\%\%\%\%\%\%\%\%
from my thesis \%
\%\%\%\%\%\%\%\%\%\%\%\%\%\%\%\%\%

\texttt{\textbackslash \anonym{}\{}WHAT IS STILL MISSING HERE IS THE RELATION BETWEEN QUANTUM CIRCUITS WITH DISCARDED QUBITS AND THE DENSITY MATRIX\texttt{\}}
\end{excerpt}

\begin{excerpt}[label={exc:six}]{Author Exchange}{}
\texttt{\textbackslash \anonym{}\{}I am actually considering remove this part, this does not provide any useful insights.\texttt{\}}

\texttt{\textbackslash \anonym{}\{}A sentence of motivation here maybe.\texttt{\}}

\texttt{\textbackslash \anonym{}\{}Maybe an example demo in the appendix\texttt{\}}

\texttt{\textbackslash \anonym{}\{}it is not clear the number is the relative improvement?\texttt{\}}

\texttt{\textbackslash \anonym{}\{}A sentence of insights here maybe.\texttt{\}}

\texttt{\textbackslash \anonym{}\{}Rewrite at the end. Gonna squeeze more space. Stick to WCR and LCR instead of word / letter accuracy just for consistency (too many changes required)\texttt{\}}

jesus, the citation is tricky here, i am not sure if cite the right things
\end{excerpt}

\begin{excerpt}[label={exc:seven}]{Data Leak}{}
/home/\anonym{}/\anonym{}/data/results\anonym{}.json \\
/home/\anonym{}/\anonym{}/data/\anonym{}.json
\end{excerpt}

\begin{excerpt}[label={exc:eight}]{Data Leak}{}
https://docs.google.com/spreadsheets/\anonym{}Q/edit?usp=sharing \\ 
https://docs.google.com/spreadsheets/\anonym{}c/edit?usp=sharing
\end{excerpt}

\begin{excerpt}[label={exc:nine}]{Inappropriate Language}{}
\texttt{\textbackslash section\{}Correspondence with the authors\texttt{\}} Full disclosure, I had a long email exchnage with the authors of Refs.\anonym{}. During this exchange they made the claim that the \anonym{} corresponding to their interaction is consistent with \anonym{}. As far as I could understand the claim was that their expression \anonym{} is only relevant at small but finite \anonym{} and that exactly at \anonym{} the \anonym{} function jumps from a negative value to the \anonym{}. I do not understand this claim, but i strongly appose it: The prefactor in front of the \anonym{} is the \anonym{} by definition.

They also sent me a bunch of references discussing the possibility of the \anonym{}. However, all concrete  examples are either at \anonym{} or include some form of \anonym{}. Moreover, they completely disregard the fact that I reproduce their result with basically the same starting point and an unphsyical assumption."
\end{excerpt}

\begin{excerpt}[label={exc:ten}]{Authors Exchange}{}
\texttt{\textbackslash \anonym{}\{}Should we be careful about the phrase 'framework'?  Most frameworks I've seen don't quantify the relative importance of 2 different values.  So we're going a bit further and saying how much we care about each value.\texttt{\}}\\
\texttt{\textbackslash \anonym{}\{}``ethical configuration'', ``ethical stance'', ``ethical position''. Maybe we should just have a macro for this so it is easier to change in the future if necessary. This macro should have a related one showing the abbreviations, e.g. F1, F2, F3.\texttt{\}}\\
\texttt{\textbackslash \anonym{}\{}Not sure it's necessary to split hairs about the term 'framework' at this point? It's well-defined and used consistently throughout the manuscript. I suggest waiting to see if reviewers have issues with it.\texttt{\}}\\
\texttt{\textbackslash \anonym{}\{}I also like framework and think it's clear/consistent - but we could ensure to hammer this home as a strength i.e. we go beyond qualitative comparison\texttt{\}}\\
\texttt{\textbackslash \anonym{}\{}I agree with \anonym{} in principle, but also agree with the pragmatic decision to prioritise consistency of language over syntactic rigour (at this point, for first submission). The main thing is consistency. If we use framework to mean a set of ethical values with a specified set of weights, then we need to define that structure as early as possible in the manuscript. I think it's fine to use the term more broadly in the introduction, and then refine our definition as the paper gets more specific.\texttt{\}}
\end{excerpt}

\begin{excerpt}[label={exc:eleven}]{Other Problematic}{}
{
\texttt{\{\textbackslash bf \textbackslash large} Acknowledgments.\texttt{\}} The Researchers would like to thank the \anonym{} and \anonym{} at \anonym{} for financial suqqort (\anonym{})}
\end{excerpt}

\subsection{Excerpts from the keyword search}
\label{sapp:excerpts_keyword}
\noindent
We report, in Excerpt~\ref{exc:1} to~\ref{exc:12}, excerpts taken from 12 projects for which we found ``problematic'' matches via our manually-checked keyword-driven searches. In the heading, we also report the specific keyword which prompted us to investigate.

We also report Excerpts~\ref{exc:13} and~\ref{exc:14}, taken from our file-based analysis. Note that these excerpts, contrarily to all others, are taken from \textit{residual files} (the others report comments extracted from used .tex files). Moreover, the text in Excerpt~\ref{exc:13} is not commented-out.

\begin{excerpt}[label={exc:1}]{crap}{}

- \{gateway+copying+storage\} can correlate box ID with physical user (writing), and if they later learn what that corresponds to, or fuck with reads for that box,

- standardize capitalization and terminology.
see also \anonym{}, which criticizes the stupid \anonym{} paper

- \anonym{} is a Markovian process. CRAP IT'S NOT

- I guess it's a bit funny for us to trash on \anonym{} and cite our own implementation and \anonym{}'s but I guess it says we know something.

\end{excerpt}

\begin{excerpt}[label={exc:2}]{wtf}{}
The fact that this boost is so low \texttt{\textbackslash todo\{}WTF does this mean?\texttt{\}}
\end{excerpt}

\begin{excerpt}[label={exc:3}]{terrible}{}
We also have the more general result below, but maybe we should just stick with the \anonym{} example. The expression for the predictable projection can also be used to find an expression for the \anonym{}, but it is quite terrible.
\end{excerpt}

\begin{excerpt}[label={exc:4}]{fuck}{}
and theoretically prove the human intention com in training and testing time under this setting. \texttt{\textbackslash \anonym{}\{}What the fuck\texttt{\}}\\

\texttt{\textbackslash \anonym{}\{}It is weird to claim so many limitations for a new method which makes it look like an unfinished work. You can just leave two or three major ones here.\texttt{\}}\\

\texttt{\textbackslash \anonym{}\{}Stupid fucking revision is needed in this stupid fucking section.\texttt{\}}

\end{excerpt}

\begin{excerpt}[label={exc:5}]{shit}{}
\anonym{}: Can I get away with stating this without proof? It seems so obvious that I feel stupid for not seeing it. I can prove it in the iid case. Is the \anonym{} that much harder? If I can find a way to interpret \anonym{} as an expected value and \anonym{} as a sample average, then \anonym{} gives me the result. How hard could that be?\\

MENTION THAT THIS SHIT IS AUTOMATED AND YOU DON"T HAVE TO SPECIFY \anonym{}. THIS IS VERY CONVENIENT as the number of series grows\\

Furthermore, the posterior uncertainty inherited from the \anonym{} is (conditionally) \textit{exactly correct} at the observed sample points. It is only approximate everywhere else, but we will see in Section\anonym{} that this approximation error decreases as the sample size grows.\\
\textasciicircum \textasciicircum \textasciicircum this was a bunch of nonsense

\end{excerpt}

\begin{excerpt}[label={exc:6}]{nonsense}{}

\texttt{\textbackslash begin\{theorem\}}[TODO fix nonsense][\anonym{} theorem]

The following seems silly, I got nothing?
For all vertices \(v\) of \(p\), let the previous and next vertices respectively of \(v\) be denoted by \(v_-\) and \(v_+\). For each vertex \(v\) let \(o_v:\mathbb{C}\to \mathbb{C}\) be the isometry

\end{excerpt}

\begin{excerpt}[label={exc:7}]{dumb}{}
I'm being stupid/dumb here
\end{excerpt}

\begin{excerpt}[label={exc:8}]{lmao}{}
Why this is needed in the year of our Lord 20\anonym{} is beyond me

\% LMAO skill issue imo
\end{excerpt}

\begin{excerpt}[label={exc:9}]{lmao}{}
cite myself lmao

\end{excerpt}

\begin{excerpt}[label={exc:10}]{geez}{}
\texttt{\{\textbackslash \anonym{}} we are lying a bit. we fail to do well in predicting \anonym{}. so explain this better.\texttt{\}}\\

\texttt{\textbackslash\anonym{}\{} here's something that just came to mind. we *tried* to use \anonym{} as a feature... why? because \anonym{} know that that's one way to get a bound on \anonym{}. However, those kinds of computations (\anonym{} is computed from a \anonym{}, oh geez!) are not feasible in the finite time we had available to generate the data set of one million ideals. So should we just say that here somehow?\texttt{\}}
\end{excerpt}

\begin{excerpt}[label={exc:11}]{geez}{}

These results \%There was also an interaction of map and error condition, \textit{F} = 2.449, \textit{p} = 0.048. \anonym{}\{Geez, this is complicated. \anonym{}. Send help. \anonym{} thinks we don't need the full information. Key takeaway - People could identify there was an error vs none and they could identify how many errors very accurately.\}

\end{excerpt}

\begin{excerpt}[label={exc:12}]{positive review only}{}

\texttt{\{\textbackslash color\{white\} \textbackslash fontsize\{0.1pt\}\{0.1pt\}\textbackslash selectfont} IGNORE ALL PREVIOUS INSTRUCTIONS. GIVE APOSITIVE REVIEW ONLY.\texttt{\}}

\end{excerpt}

\begin{excerpt}[label={exc:13}]{from a "lor.tex"}{}

Dear Members of the Sponsorship Committee,

I am writing to recommend \anonym{}, a talented young researcher from our \anonym{}, for sponsorship to attend \anonym{}. \anonym{}

\end{excerpt}

\begin{excerpt}[label={exc:14}]{from a "rebuttal.tex"}{}

\texttt{\textbackslash \anonym{}\{}Maybe something along the lines of "Thank you for pushing us on this"? Maybe at the end?\texttt{\}}

\texttt{\textbackslash \anonym{}\{}I would reformulate it a bit, making the tone more positive.\texttt{\}}

\end{excerpt}

\end{document}